# Polar order in a fluid like ferroelectric with a tilted lamellar structure – observation of a polar smectic C ($SmC_P$) phase

Jordan Hobbs*[1], Calum J. Gibb[2], Damian Pociecha[3], Jadwiga Szydłowska[3], Ewa Górecka[3], and Richard. J. Mandle [1,2]

[1] School of Physics and Astronomy, University of Leeds, Leeds, UK, LS2 9JT
[2] School of Chemistry, University of Leeds, Leeds, UK, LS2 9JT
[3] Faculty of Chemistry, University of Warsaw, ul. Zwirki i Wigury 101,02-089, Warsaw, Poland

*Author for correspondence e-mail: j.l.hobbs@leeds.ac.uk

## Abstract

The discovery of fluid states of matter with spontaneous bulk polar order is appreciated as a major discovery in the fields of soft matter and liquid crystals. Typically, this manifests as polar order superimposed atop conventional phase structures and is thus far limited to orthogonal phase types. Here we report a family of materials which exhibit a previously unseen state of matter which we conclude is a polar smectic C phase, and so we term it $SmC_P$. The spontaneous polarisation of the $SmC_P$ phase is over two orders of magnitude larger than that found in conventional ferroelectric SmC phase of chiral materials used in some LCD devices. Fully atomistic molecular dynamics simulations faithfully and spontaneously reproduce the proposed structure and associated bulk properties; comparison of experimental and simulated X-ray scattering patterns shows excellent agreement. The materials disclosed here have significantly smaller dipole moments than typical polar liquid crystals such as RM734 which suggests the role of molecular electrical polarity in generating polar order is perhaps overstated, a view supported by consideration of other molecular systems.

## Introduction

Since its discovery in 2017 [1–3] and subsequent confirmation of discovery in 2020 [4–7], the ferroelectric nematic phase ($N_F$) has been shown to have a rich variety of properties such as non-linear optical properties [8], extreme interfacial instabilities [9], bulk photovoltaic effect [10], extreme polarisation screening [11], and piezoelectric properties [12] to name a few, where all these properties have their origins in the large longitudinal spontaneous polarisation (> 4 μC $cm^{-2}$) caused by the lack of inversion symmetry of the $N_F$ phase (**Fig. 1a**) [13,14].

Longitudinal polarisation of liquid crystalline materials is not isolated to the nematic phase and more exotic phase structures have been discovered such as the ferroelectric smectic A ($SmA_F$) phase [15–17] where the polarisation vector is parallel to the layer normal. An antiferroelectric smectic A ($SmA_{AF}$) phase has also recently been discovered [18,19] although the exact nature of the anti-ferroelectric ordering within those systems is not yet clear. Perhaps even more interesting is the recent discovery of longitudinally polar LC phases which spontaneously break chiral symmetry, presumably due to the high spontaneous polarisation, such as the so-called twist bend ferroelectric nematic phase ($N_{TBF}$) [20,21] and the heliconical polar smectic C phase ($SmC_H^P$) [18]. It has recently been suggested that since these LC phases with longitudinal polarity are proper ferroelectrics (i.e. caused by dipolar interactions rather than steric interactions) and so polar order is effectively superimposed over the orientation and positional order of the underlying LC phase [19]. A logical progression of this idea would suggest the existence of longitudinally polar variants of many of the known non-polar LC phases.

In apolar liquid crystals, it is widely understood that fluorination of the rigid core unit generally suppresses smectic phase formation [22,23] and it was shown that removal of adjunct fluorine atoms from the archetypal ferroelectric nematogen **DIO** [3] leads to the formation of a $SmA_F$ phase [16]. The dioxane moiety found in **DIO** is also known to supress smectic phase formation [24]. Here we show that exchange of the dioxane ring for a phenyl moiety coupled with the removal of all adjunct fluorine atoms (the **PPZGU-*n*-F** series**, Fig. 1b)** further promotes the formation of smectic phases with bulk polar order, leading to the observation of a both a $SmA_F$ phase and a polar, titled smectic phase (**Fig. 1a [right]**).

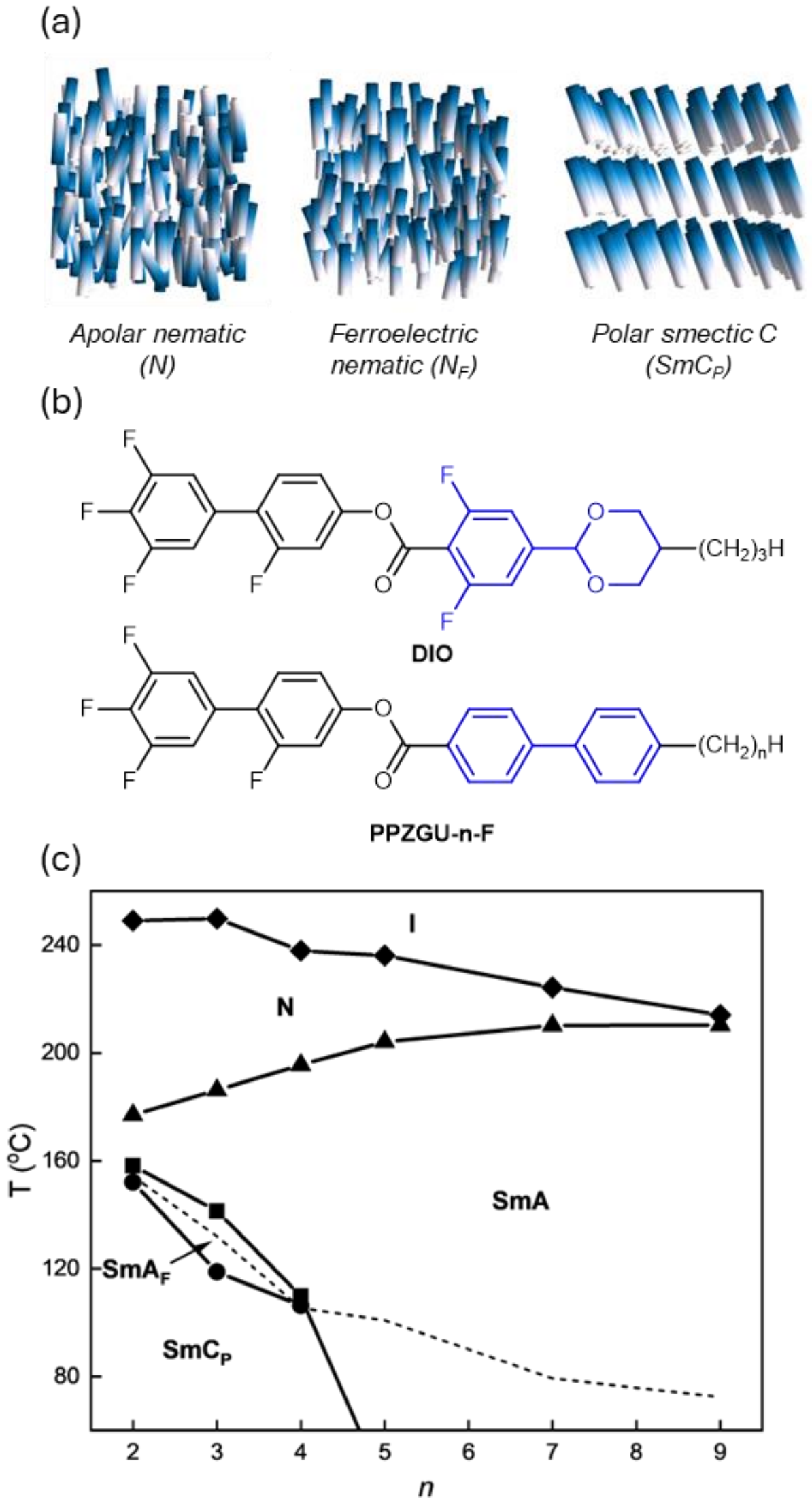

**Fig. 1.** Schematic representations of the **(a)** conventional N, $N_F$ and $SmC_P$ mesophases; **(b)** the chemical structure of the archetypal $N_F$ materials **DIO** [3] and the general structure of the **PPZGU-*n*-F** materials reported in this work.

The difference in molecular structure between **DIO** and **PPZGU-n-F** are highlighted in blue; **(c)** phase diagram showing the evolution of phase behaviour on increasing alkyl chain length (***n***) for the **PPZGU-*n*-F** materials. Melting points are denoted by the dotted line.

## Results and Discussion

Transition temperatures of the **PPZGU-*n*-F** materials are given in **Figure 1d** (tabulated data in the ESI; **Table S1**) and were determined by differential scanning calorimetry (DSC), with phase assignments made by polarised optical microscopy (POM), assisted by X-ray scattering and current response studies (*vide infra*). For the sake of brevity, we will focus our discussion on the homologue with *n*=3 as a representative example of the phase behaviour of the **PPZGU-*n*-F** series with supplementary results for the other materials found in the ESI.

**PPZGU-3-F** shows apolar nematic (N) and smectic A (SmA) phases (as do all other homologues), identified using POM and X-ray scattering, as well as two further LC phases observed on further cooling. Confinement between untreated glass slides typically resulted in homeotropic alignment for the N and SmA phases while for the two lower temperature phases, uncharacteristic textures were observed. However, confinement in cells treated for homeotropic alignment resulted instead in degenerate planar alignment (**Fig. S1**). In X-ray scattering experiments, both the N and SmA phases show diffuse scattering at wide-angles, indicating liquid-like order in direction perpendicular to the director. The lamellar structure of the SmA phase leads to the Bragg scattering of X-rays at small angles (whereas the nematic has only diffuse scattering), yielding a layer spacing (*d*) of 25.0 Å (**Fig. 2a**) which is largely temperature independent across the entire phase range. To obtain the layer spacing (*d*) as a ratio of molecular length (*L*) we performed molecular geometry optimisation at the B3LYP-GD3BJ/cc-pVTZ level; using this value affords a ratio of ~1.07, indicating a near monolayer arrangement of the molecules within the smectic layers. Current response measurements demonstrate that the N and SmA phases are apolar, with only small ionic and dielectric contributions being detected. Notably however, below 150 °C (still in the SmA phase), a single peak centred around polarity reversal is observed in the SmA phase (**Fig. S4**), presumably the result of pre-transitional field induced polarisation.

For the three shortest homologues studied (*n*=2-4), two further phases are observed upon cooling the apolar SmA phase. X-ray scattering shows that both phases have a liquid-like smectic structure, with diffuse scattering at wide angles and sharp Bragg scattering in the small angle region (**Fig. 2b**). Subsequent experiments, detailed below, identify these phases as $SmA_F$ and a newly discovered tilted phase with bulk polar order. At the transition from the SmA into the $SmA_F$ phase (i.e. the onset of polar order signified by a single peak in the current response, **Fig. 2c**) there is a slight contraction of layer spacing while the low-angle peak remains a single lobe. From quantum chemical calculations at the B3LYP-GD3BJ/cc-pVTZ level of DFT **PPZGU-3-F** is shown to have an angle of 10.7 ° between the dipole vector and the long molecular axis (**Fig. 2d**, **Table S2**). A $SmA_F$ phase comprised of such molecules could result in a layered structure where the dipoles are aligned normal to the layer planes with the long molecular axes inclined at an angle relative to the layer normal. For this phase to be uniaxial, and be SmA type rather than SmC, the inclined molecules would need to be distributed around a cone and such a structure would be analogous to a de Vries SmA structure[25]. Such a structure would explain the reduction in layer spacing (**Fig. 2a**) in the $SmA_F$ phase. We also note for the Bragg scattering associated with the lamellar structure up to $6^{th}$ order harmonic can be observed, although this limit is due to the Q limits of the detector rather than because we could not observe the $7^{th}$ order fringe (**Fig. S3**). Generally, for a SmA phase the layers are diffuse and more of a density modulation rather than strict layers. However, the observation here of several harmonics of the diffraction signal due to layered structure means that the density modulation along layer normal is strongly non-sinusoidal giving rather sharp boundaries between the high- and low-density regions. The long-range correlation of the layers may be due to a lack of layer undulations as these would lead to areas of bound charge due to splay of the polarisation in those regions[26].

For the $SmA_F$ phase the current response shows that the pretransition peak observed for the SmA phase moves to higher voltage, indicating the increased threshold for switching (**Fig. 2c**). This increase is possibly due to the increased cost of rotating the smectic layers along with the polarisation where the maximum value of polarisation in the $SmA_F$ phase is ~3.0 μC $cm^{-2}$ (**Fig. S5**). The non-centrosymmetric structure of the $SmA_F$ phase is further affirmed by second harmonic generation (SHG) microscopy (**Fig. S6**), where the presence of an SHG signal without biasing electric field confirms the polar ground state of the $SmA_F$ phase.

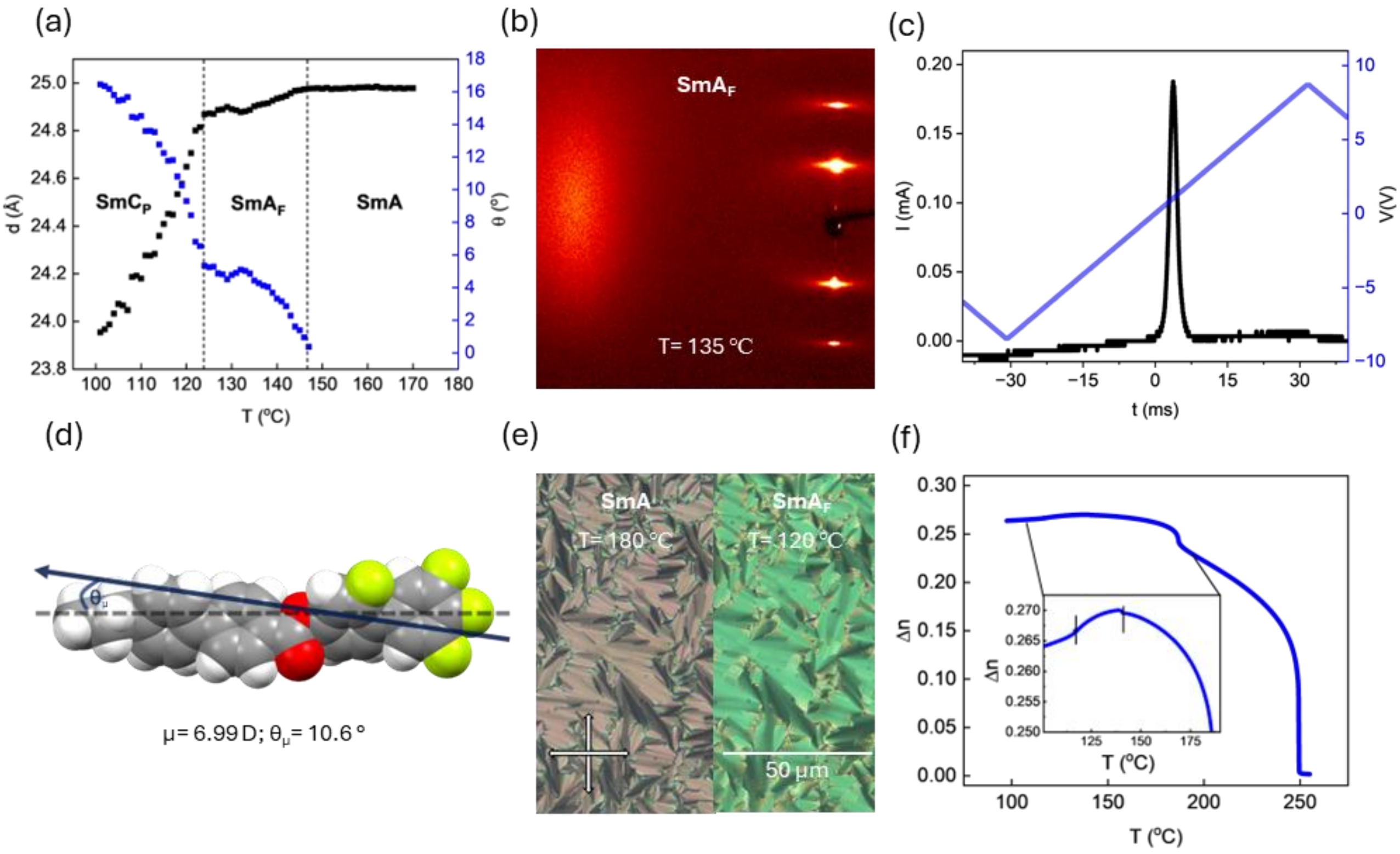

**Fig. 2.** **(a)** Temperature dependence of the layer spacing (d) for **PPZGU-3-F** in the SmA, $SmA_F$ and $SmC_P$ phases, **(b)** 2D x-ray diffraction pattern for **PPZGU-3-F** in the $SmA_F$ phase (T= 135 ℃), (c) current response trace measured at 8 Hz in the $SmA_F$ phase (T = 136 ℃), **(d)** the energy minimised geometry of **PPZGU-3-F** calculated at the B3LYP-GD3BJ/cc-pVTZ level of DFT. The long molecular axis is marked by the hashed line with the direction of the longitudinal molecular dipole moment (inclined at an angle relative to the long molecular axis) marked by the arrow (μ= 6.99 D; θμ= 10.6 °), (e) POM micrographs of the SmA (T= 180 ℃) [left] and $SmA_F$ [right] (T= 120 ℃) phases. **PPZGU-3-F** samples are confined in thin cells treated for a homeotropic anchoring condition, and (f) the temperature dependence of the optical birefringence, T(Δ*n*), for **PPZGU-3-F** measured using green light, λ= 532 nm in a 1.7 μm cell treated for planar alignment.

Optically, the $SmA_F$ phase retains the fan texture observed in the proceeding apolar SmA phase, however the texture now contains regions of rapidly changing retardance (**Fig. S1**). These defects generally form out of fan domain boundaries of the SmA phase, indicative of the rotation of the vertical director at regions where the sign of the polarization switches without forming a line defect (**Fig. 2e**). Such observations are suggested to be characteristic of the $SmA_F$ phase [18,19]. We propose that the texture observed in these materials is somewhat paramorphotic, as it has been suggested that polar ordering prohibits the formation of Dupin cyclides responsible for the fan and focal conic textures in the apolar SmA phase [16]. In anti-parallel rubbed planar cells, a clear transition can only be observed when comparing the dark states of both phases via comparison of the defects which form around the spacer beads (**Fig. S7**). In general, when anti-parallel rubbed planar cells are used to confine the $N_F$ phase, π-twist domains can form, due to polar surface interactions [27] and to minimize the bulk polarisation of the system [28]. Such domains are not observed within the $SmA_F$ phases studied here, consistent for when the phase is formed from a preceding $N_F$ phase [15]. Optical birefringence ($\Delta n$) was measured by cooling throughout the whole LC phase range (**Fig. 2f**). In the N phase a power law temperature dependence was observed, with a small jump in $\Delta n$ on transition to the SmA phase. Another slight step (~0.01) was detected for the SmA-$SmA_F$ transition before a slight decrease is observed throughout the $SmA_F$ phase. Although this could be due to a slight reduction in alignment quality, POM images of the dark state of these materials does not imply any change in the quality of alignment and we instead suggest that this is further evidence for the adoption of a de Vries like molecular orientation.

Further cooling of the $SmA_F$ phase for homologues with $n < 5$ sees a transition to a final liquid-like smectic phase. Here, the Bragg scattering of the $SmA_F$ is retained though a continuous decrease in the layer spacing, indicating a tilting of the molecules away from the layer normal (i.e. a SmC-like phase), is now observed. For all the homologues studied, the tilt does not saturate at the point of crystallisation (~ 16 ° at 100 °C for **PPZGU-3-F**; **Fig. 2a**). Within a cell treated for homeotropic anchoring, the fan-like texture of the preceding $SmA_F$ splits into small domains, forming a broken fan-like texture (**Fig. 3a**), where each domain corresponds to different tilt direction. The formation of these domains disrupts the alignment of the sample when confined within cells treated for planar anchoring, resulting in the complete loss of any extinction condition upon rotating the cell with respect to the polarisers' direction (**Fig. S8d**). The lack of alignment in planar cells prohibits the reliable measurement of birefringence via the technique used here. Measurements of the current response reveal that the polarity of the $SmA_F$ phase is retained (**Fig. 3b**), with the position of the main peak observed in the preceding phase moving backwards towards lower voltage indicating a decreased elastic cost of bulk polarisation reversal, with the $P_S$ saturating at a value of 3.3 $\mu C\ cm^{-2}$. However, a second peak of a smaller magnitude to the main signal, emerges in the current response upon entering the new phase just before applied voltage polarity reversal. Such a current response would not result in a truly "ferroelectric" hysteresis loop, owing to the observation of multiple peaks. This peak is possibly due to some ionic effect induced by the emergence of tilt domains, or it could be associated with tilt removal, analogous to a mechanism suggested for the recently discovered helical polar SmC phase [18], though we note no evidence for helix formation is observed for the materials presented here. The polar nature of the lowest temperature smectic phase is further confirmed by SHG microscopy (**Fig. S6**) though whether the phase is "ferroelectric" or some other type of polar ordering is not yet clear from performed measurements and thus we term this phase the polar smectic C phase ($SmC_P$).

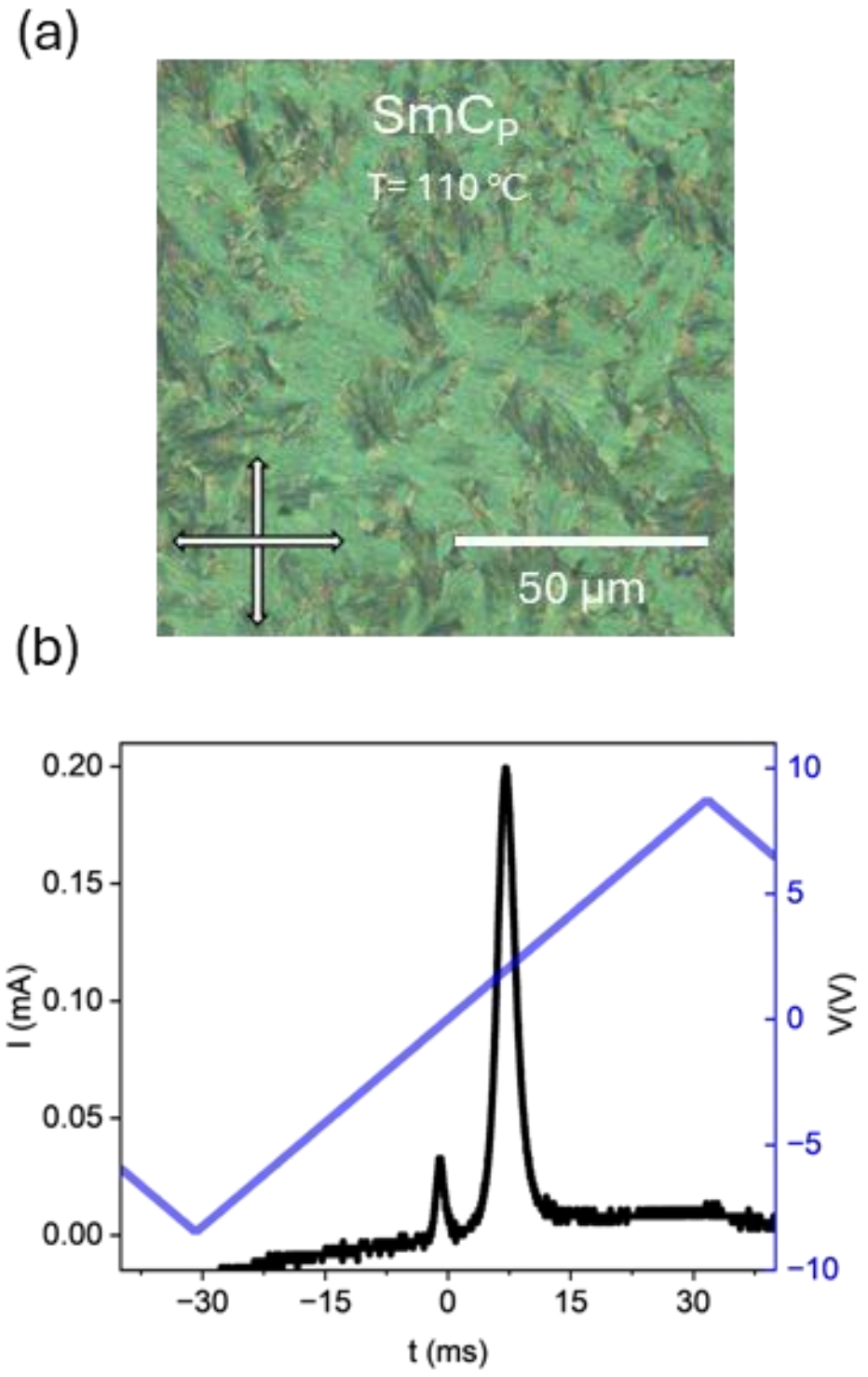

**Fig. 3.** **(a)** POM micrographs of the broken fan texture of the $SmC_P$ (T= 110 ℃) phase observed for **PPZGU-3-F**. The image is taken of a sample confined within a thin cell treated for homeotropic anchoring; **(b)** current response trace measured at 8 Hz in the $SmC_P$ phase (T = 104 ℃) showing the emergence of a second smaller peak at longer time scales.

We sought to validate the structure proposed above using fully atomistic molecular dynamics simulations, during which (and under certain conditions) the $SmC_P$ phase forms spontaneously. Our simulations initially begin from either an isotropic or polar nematic configuration, which evolve into apolar and polar phases, respectively. Gratifyingly, both apolar and polar initial configurations yield the correct phase sequences, with bulk properties and transition temperatures in reasonable agreement with experiment. Simulations commencing from an isotropic state remain isotropic at and above 252 °C, while the simulation at 227 °C evolves into a nematic phase ($\langle P2 \rangle$ = 0.61 +/- 0.04) (**Fig. S8**). At and below 202 °C, the simulations evolve into an SmA phase with layer spacing of 2.4 +/- 0.1 nm (**Fig. S10**). At no point does the apolar simulation evolve into a tilted smectic phase.

Simulations commencing from a polar nematic configuration evolve into an isotropic state (at and above 252 °C) and slowly transform to a non-polar N phase on cooling (227 °C). The orientational order parameter is retained, while the polar order parameter tends to zero over the first 140 ns (**Fig. S9**). At and below 202 °C, a polar SmA phase spontaneously forms with $\langle P1 \rangle$ = 0.94 +/- 0.01 and $\langle P2 \rangle$ = 0.78 +/- 0.02. The simulated SmA phase has a layer spacing of 2.3 (+/- 0.1) nm and spontaneous polarisation value of 2.5 (+/- 0.02) μC $cm^{-2}$, both of which are in harmony with the values obtained by experiment. Crucially, at and below 202 °C, the molecules spontaneously tilt away from the layer normal, resulting in the formation of a tilted smectic phase with bulk polar order, analogous to the $SmC_P$ phase proposed based on experimental observations (**Fig. 4c**). The *in-silico* $SmC_P$ phase has a layer spacing of 2.2 +/- 0.1 nm and a tilt angle of 25 +/- 3 °, which are in reasonable agreement with experimentally obtained values (2.4 nm and 16 °, respectively). The $SmC_P$ phase has a near saturated polar order parameter (<$P$1> = 0.90 +/- 0.04) and orientational order consistent with a fluid smectic phase (<$P$2> = 0.78 +/- 0.06).

We used the MD trajectory to simulate a 2D SAXS pattern of the $SmC_P$ phase by mapping the atomic electron density onto a 3D grid and transforming into the structure factor via Fourier transform (see ESI). The simulated 2D SAXS pattern closely matches that obtained experimentally (**Fig. 4a** and **4b**), with diffuse wide-angle scattering and multiple high intensity low-angle peaks. Periodic noise in the simulated SAXS background arises from the Fourier Transform procedure. Additionally, the system size (1000 molecules) and limited simulation duration (500 ns) conspire to introduce further artifacts, such as finite-size effects and incomplete sampling of long-range correlations. The qualitative agreement between experimental/simulated 2D SAXS patterns, along with the ability of the MD simulation to accurately recreate experimentally measured properties, gives a high degree of confidence in our model of the $SmC_P$ phase.

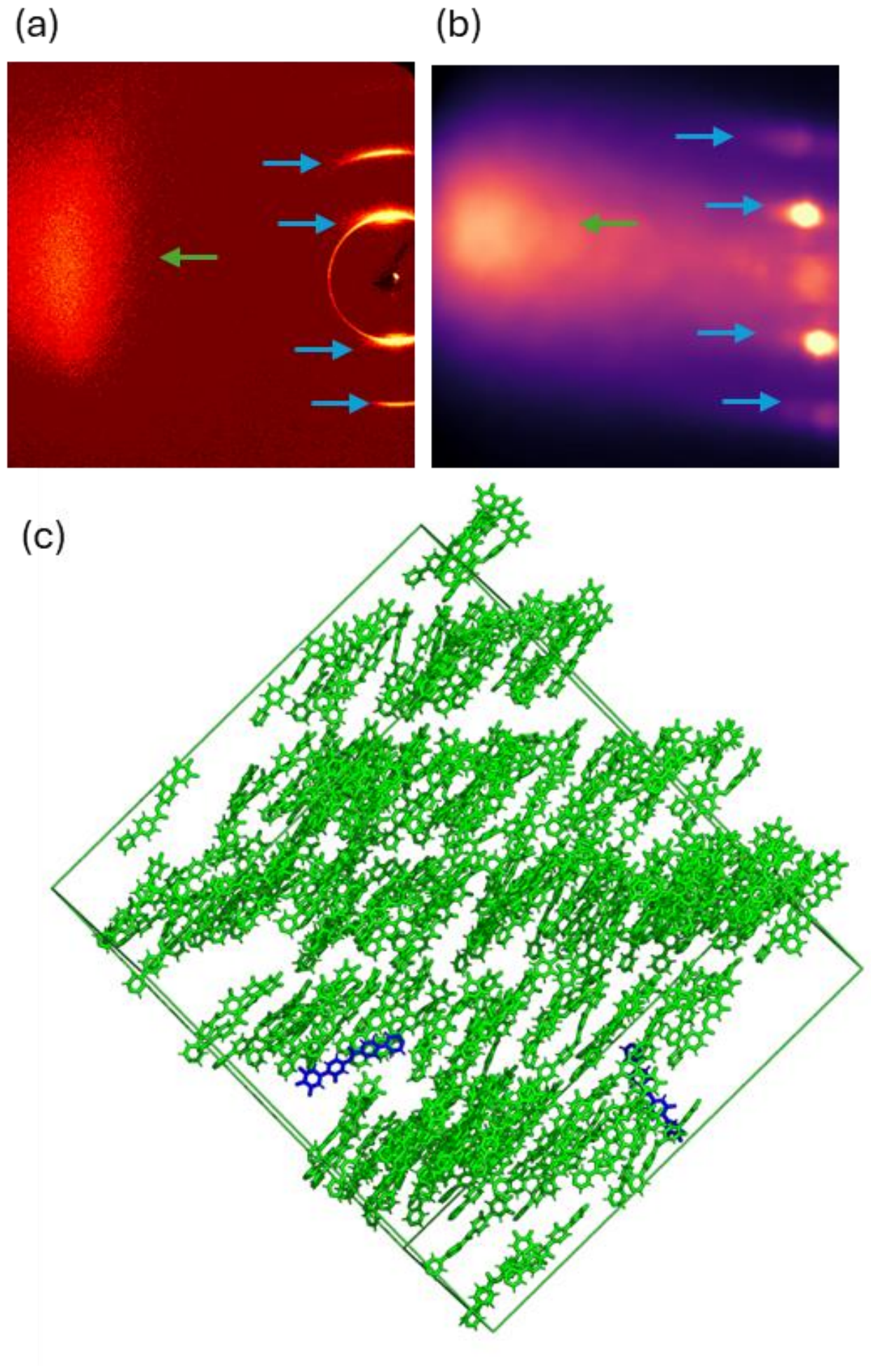

**Fig. 4** **(a)** an experimental 2D x-ray diffraction pattern for **PPZGU-3-F** in the $SmC_P$ phase (T= 115 ℃); **(b)** simulated 2D SAXS pattern of the $SmC_P$ phase of **PPZGU-3-F** at 127 °C based on the MD trajectory showing good agreement with the data obtained by experiment; and **(c)** Instantaneous configuration of the *polar* smectic C phase for the ***n*=3** homologue formed spontaneously from the initial polar configuration at the same temperature (<$P$2> = 0.78 +/- 0.06). The snapshot is color-coded green/blue to show molecules whose dipole is parallel or antiparallel with the director, respectively (<$P$1> = 0.90 +/- 0.04). The hydrocarbon chains are omitted from the rendering to aid visualisation of the layer structure (layer spacing = 2.3 +/- 0.1 nm; tilt angle = 25 +/- 3 °). The calculated spontaneous polarisation ($P_s$) is 3.2 +/- 0.01 μC.cm-$^2$. Only 400 molecules (out of 1000) are shown, to aid visualisation.

Although the interpretation of the dielectrics of longitudinally polar liquid crystal phases is not yet fully understood [29–34], we elected to probe the dynamic properties of **PPZGU-3-F** by dielectric spectroscopy in the SmA, $SmA_F$ and $SmC_P$ phases. Complex permittivity data was obtained for samples in 10 μm thick cells with untreated gold electrodes (**Fig. S12a, S12b**), with the caveat that the dielectric strengths and relaxation times of any relaxation processes will be dependent on the sample thickness, orientation, and anchoring behaviour. **Figure 5a** contains representative dielectric loss data for **PPZGU-3-F** for the $SmA_F$ and $SmC_P$ phases (details of fitting given in ESI with example of the deconvoluted fits, and **Figures 5b and 5c** show the relaxation times and dielectric strength for the processes observed through the SmA, $SmA_F$ and $SmC_P$ phases (α fitting parameter is shown in **Fig. S12c**). In the SmA phase, only a single dielectric process was observed (P1). Power laws were used to account for a low frequency conductivity flank and for any processes observed at timescales faster than the frequency window. P1 shows a super-Arrhenius, Vogel-Fulcher-Tammann (VFT) dependence of its relaxation times typical of cooperative processes with asymptotic behaviour at the transition to the $SmA_F$ phase. We suggest that this process is related to rotations around the short molecular axis which become frozen out at the SmA-$SmA_F$ transitions due to the emergence of polar order. Due to the formation of small ferroelectric clusters in the SmA phase, indicated by the strong pre-transitional induced polarisation (**Fig. S5**), the dielectric strength of P1 also shows significant growth approaching the transition to the polar phase (~100 at the SmA-$SmA_F$ transition), behaviour analogous to the $N_F$ phase [35].

In the $SmA_F$ phase two processes are observed. A weak mode (P2) appears with a timescale of roughly $10^{-5}$ s and very large distribution parameter. The timescales and dielectric strength of this process are essentially temperature independent in the $SmA_F$ phase, only consistently moving to longer timescales and steeply increasing in strength (circa 2 orders of magnitude) upon approach to the $SmC_P$ phase. The main relaxation mode observed in the $SmA_F$ phase (P3) is moving into the frequency window from shorter timescales. As the $SmC_P$ phase is approached, P3 steeply increases in strength and slows down (Fig. 5b, c). In the $SmC_P$ phase, the strong P3 dominates dielectric response, it is only weakly temperature dependent for both its timescale and dielectric strength (in fitting procedure the much weaker P2 mode was also used to ensure the continuous evolution of the modes). We suggest that P3 that starts to develop in $SmA_F$ phase is related to tilt fluctuations, which increase in strength as the molecular motions become more cooperative near phase transition to tilted $SmC_P$ phase. In $SmC_P$ phase the relaxation process P3 is attributed to collective rotation of molecules, and thus polarization, on the tilt cone.

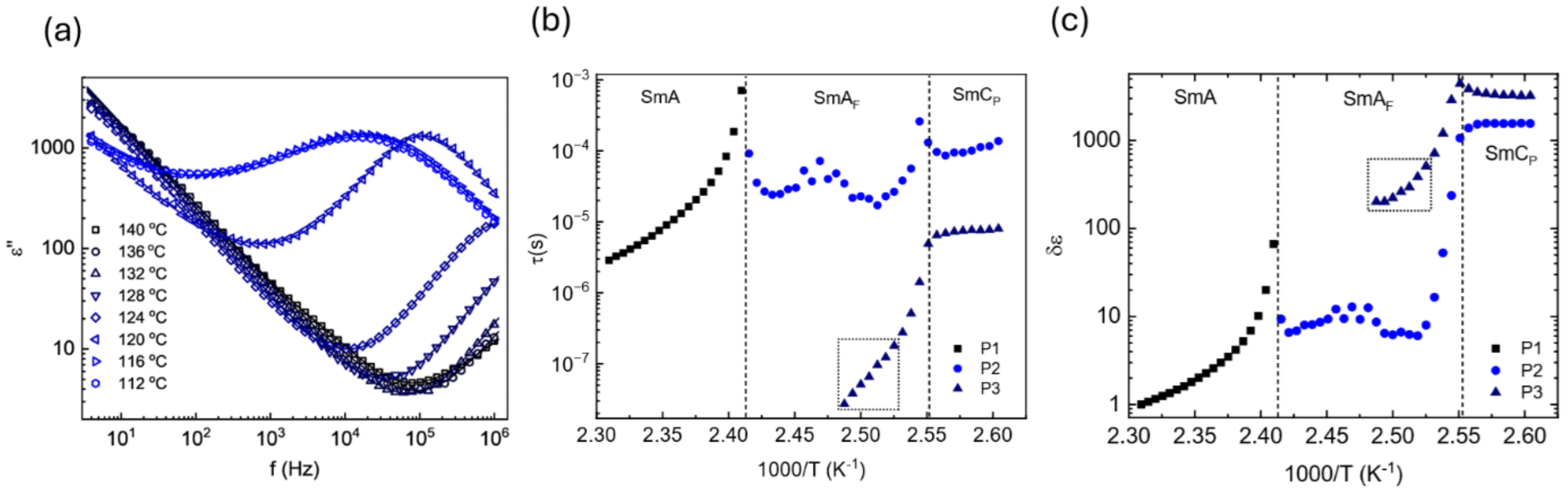

**Fig. 5** **(a)** Representative dielectric loss spectra for **PPZGU-3-F** in the SmA$_F$ and SmC$_P$ phases. Line through the data corresponds to the fits obtained; **(b)** Arrhenius plot of relaxation times; and **(c)** dielectric strength parameter obtained. The dashed box area in each plot corresponds to areas where P3 was at the edge and outside the fitting window reducing the fit quality in that region.

Ultimately the generation of polar ordering in real-world experimental conditions is principally a product of molecular structure, and while a full structure-property relationship is beyond the scope of this initial communication, some striking trends are immediately apparent. Firstly, considering **DIO** [3] as a starting point, we see that the removing of fluorine atoms from the molecular structure gives a progressively higher onset temperature for polar ordering (either to N$_F$, N$_X$, or SmA$_F$; **Fig. 6a**)[16]. Similarly, replacement of the dioxane unit of **DIO** with phenyl ring - material **PIO** [19] - increases the propensity for smectic phase formation (**Fig. 6b**). Further reduction of the molecular polarity, achieved by removing fluorine substituents from **PIO**, yields the **PPZGU-*n*-F** materials reported here (**Fig. 6b**) which dramatically increases the onset temperature of polar ordering. This runs counter to the view that molecular polarity alone generates bulk polar ordering, evidenced by calculated molecular dipole moments which are given in **Figure 6b,** and we note that the molecules presented here result in one of the lowest reported molecule dipoles to form an LC phase [36]. Even further reduction in polarity by removing subsequent fluorine atoms gives non-polar phase types (**PPZPU-3-F**; this work, see ESI). We also note the importance of molecular length in arresting head-to-tail flipping, as evidenced by **PZGU-3-F** [19]. **(Fig. 6c)**, which is non-polar. Generally, as the degree of fluorination decreases the propensity for smectic phase formation increases, which has led us to the discovery of the SmC$_P$ phase.

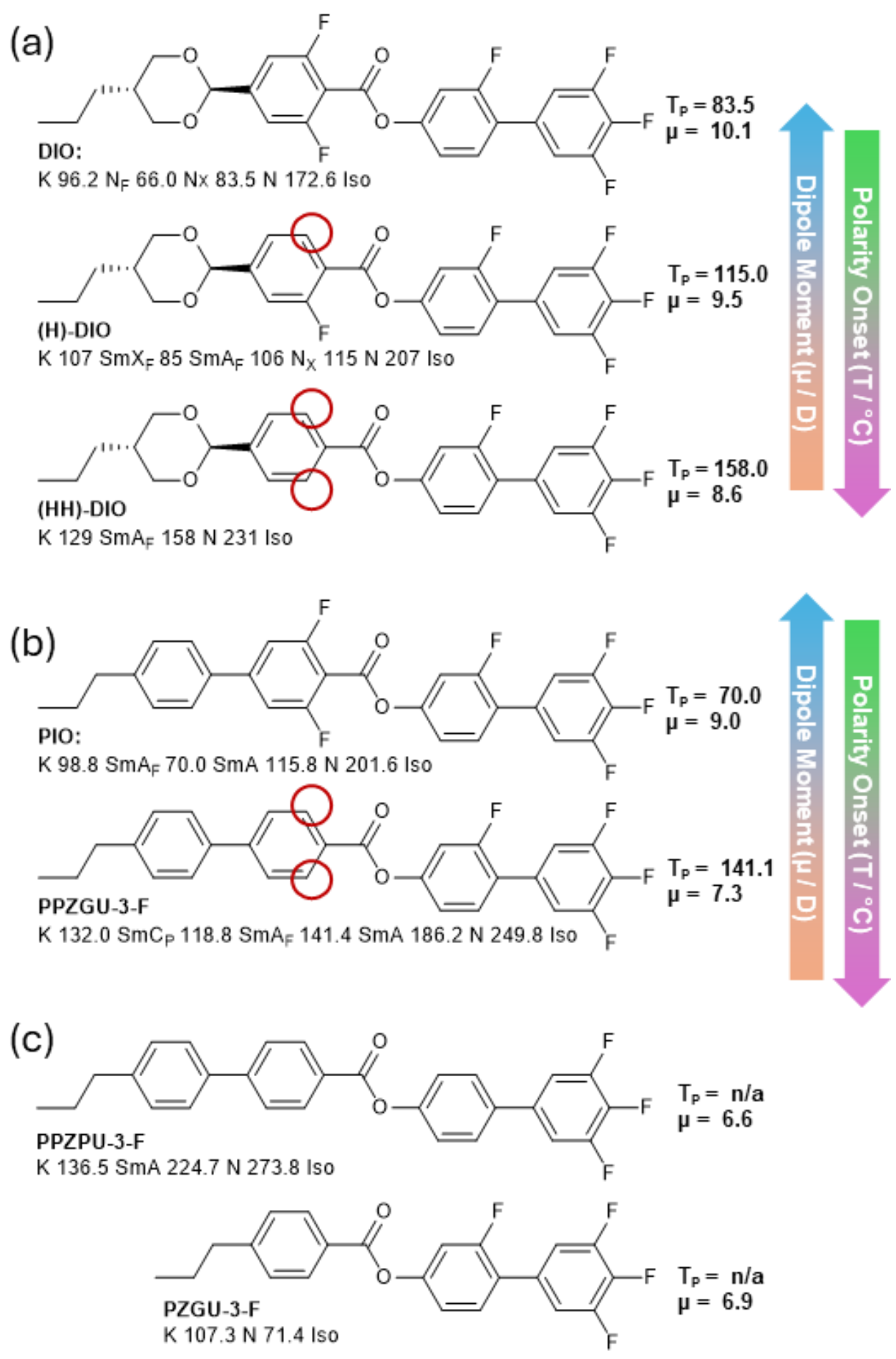

**Fig. 6.** Molecular structures, transition temperatures (°C) and dipole moments (μ; at the B3LYP-GD3BJ/cc-pVTZ level of DFT) of **(a) DIO** [3] and similar selected analogues; [16] **(b) PIO** [19]and **PPZGU-3-F**, the later from this work; **(c)** the non-polar derivatives of **PPZGU-3-F** and **PIO**, **PPZPU-3-F** (of this work) and **PZGU-3-F** [19], respectively.

## Conclusions

We have presented evidence for a new class of polar ordered materials with a titled lamellar structure which we conclude to be an analogue of the traditional SmC phase, which we term the $SmC_P$ phase. The $SmC_P$ phase presents as a conventional tilted smectic phase with superimposed polar order with the direction of polarisation being along the director and tilted away from the layer normal. This polar SmC phase occurs without helix formation as was reported recently [18], and similar phase sequence has been seen recently reported by others [37,38]. In the same materials we also observe a polar orthogonal smectic phase ($SmA_F$). The observation of polar ordered phases in the structures presented here is surprising given their rather modest polarity, with dipole moments comparable to 5CB. Both the $SmA_F$ and the $SmC_P$ phase emerge spontaneously in MD simulations, further highlighting the power of this simulation technique as a computational microscope for the investigation of polar liquid crystalline systems.

An important observation regarding the materials presented here is that polar order is only seen in shorter homologues (***n*** < 5) which is likely a consequence of the increased physical separation mandated by long terminal chains disrupting the head-to-tail packing of molecules. Empirically, this reinforces the rather general observation of the importance of head-to-tail correlations between molecules in the formation of polar LC phase [4,39].

### Data Availability

The data associated with this paper are openly available from the University of Leeds Data Repository at https://doi.org/10.5518/1573.

### Acknowledgements

R.J.M. thanks UKRI for funding *via* a Future Leaders Fellowship, grant no. MR/W006391/1, and the University of Leeds for funding *via* a University Academic Fellowship. E.G. and D.P. thanks the National Science Centre (Poland), grant no. 2021/43/B/ST5/00240. R.J.M. acknowledges ongoing support from Merck KGaA. Computational work was performed on the ARC4 computer, part of the high-performance computing resource available at the University of Leeds.

### Author Contributions

J.H. and C.J.G contributed equally to this work; C.J.G. and R.J.M; performed chemical synthesis; J.H. and C.J.G. performed microscopy and birefringence studies; J.H. performed applied field studies and DSC analysis; D.P performed X-ray scattering experiments, assisted by J. H and C.J.G; D.P. performed dielectric spectroscopy which was analysed by J.H; J.S. performed SHG microscopy measurements; J.H. and R.J.M. performed electronic structure calculations; R.J.M performed molecular dynamics simulations; R.J.M. and D.P secured funding. The initial manuscript was written jointly by J.H, C.J.G, and R.J.M before being subsequently reviewed with contributions from all authors.

## Supplementary information

# Polar order in a fluid like ferroelectric with a titled lamellar structure – observation of a polar smectic C ($SmC_P$) phase

Jordan Hobbs*[1], Calum J. Gibb[2], Damian Pociecha[3], Jadwiga Szydłowska[3], Ewa Gorecka[3], and Richard. J. Mandle[1,2]

[1] School of Physics and Astronomy, University of Leeds, Leeds, UK, LS2 9JT

[2] School of Chemistry, University of Leeds, Leeds, UK, LS2 9JT

[3] University of Warsaw, Faculty of Chemistry, ul. Zwirki i Wigury 101,02-089Warsaw, Poland

*Author for correspondence e-mail: j.l.hobbs@leeds.ac.uk

**Contents**

# 1 Supplementary Methods

## 1.1. Chemical Synthesis

Chemicals were purchased from commercial suppliers (Fluorochem, Merck, ChemScene, Ambeed) and used as received. Solvents were purchased from Merck and used without further purification. Reactions were performed in standard laboratory glassware at ambient temperature and atmosphere and were monitored by TLC with an appropriate eluent and visualised with 254 nm light. Chromatographic purification was performed using a Combiflash NextGen 300+ System (Teledyne Isco) with a silica gel stationary phase and a hexane/ethyl acetate gradient as the mobile phase, with detection made in the 200-800 nm range. Chromatographed materials were filtered through 200 nm PTFE frits and then subjected to re-crystallisation from an appropriate solvent system. Yields refer to chromatographically and spectroscopically homogenous material.

## 1.2. Chemical Characterisation

Chemical materials were characterised by NMR spectroscopy using either a Bruker Avance III HDNMR spectrometer operating at 400 MHz, 100.5 MHz or 376.4 MHz ($^{1}$H, $^{13}$C{$^{1}$H} and $^{19}$F, respectively) or a Bruker AV4 NEO 11.75T spectrometer operating at 500 MHz, 125.5 MHz or 470.5 MHz ($^{1}$H, $^{13}$C{$^{1}$H}).

## 1.3. Phase characterisation

Phase transition temperatures and associated enthalpies of transition were determined by differential scanning calorimetry (DSC) using a TA instruments Q2000 heat flux calorimeter with a liquid nitrogen cooling system for temperature control. Samples were measured with 10 ℃ min$^{-1}$ heating and cooling rates. The transition temperatures and enthalpy values reported are averages obtained for duplicate runs. Phase transition temperatures were measured on cooling cycles for consistency between monotropic and enantiotropic phase transitions, while crystal melts were obtained on heating. Phase identification by polarised optical microscopy (POM) was performed using a Leica DM 2700 P polarised optical microscope equipped with a Linkam TMS 92 heating stage. Samples were studied sandwiched between two untreated glass coverslips.

### 1.4. Measurement of Spontaneous Polarization ($P_S$)

Spontaneous polarisation measurements are undertaken using the current reversal technique [40,41]. Triangular waveform AC voltages are applied to the sample cells with an Agilent 33220A signal generator (Keysight Technologies), and the resulting current outflow is passed through a current-to-voltage amplifier and recorded on a RIGOL DHO4204 high-resolution oscilloscope (Telonic Instruments Ltd, UK). Heating and cooling of the samples during these measurements is achieved with an Instec HCS402 hot stage controlled to 10 mK stability by an Instec mK1000 temperature controller. The LC samples are held in 4µm thick cells with no alignment layer, supplied by Instec. The measurements consist of cooling the sample at a rate of 1 K min$^{-1}$ and applying a triangular voltage wave at a frequency of 8 Hz using 10 $V_{RMS}$.

There are three contributions to the measured current trace: accumulation of charge in the cell ($I_c$), ion flow ($I_i$), and the current flow due to polarisation reversal ($I_p$). To obtain a $P_S$ value, we extract the latter, which manifests as one or multiple peaks in the current flow, and integrate as:

$$P_S = \int \frac{I_p}{2A} dt \qquad \textbf{(2)}$$

where A is the active electrode area of the sample cell.

### 1.5. X-ray measurements

X-ray diffraction measurements of samples in liquid crystalline phases were carried out using a Bruker D8 GADDS system, equipped with micro-focus-type X-ray source with Cu anode and dedicated optics and VANTEC2000 area detector. Small angle diffraction experiments were performed on a Bruker Nanostar system (IµS microfocus source with copper target, MRI heating stage, Vantec 2000 area detector).

### 1.6. Optical Birefringence

Optical birefringence was measured with a setup based on a photoelastic modulator (PEM-90, Hinds) working at a modulation frequency f = 50 kHz; as a light source a halogen lamp (Hamamatsu LC8) equipped with narrow bandpass filters was used. The transmitted light intensity was monitored with a photodiode (FLC Electronics PIN20) and the signal was deconvoluted with a lock-in amplifier (EG&G 7265) into 1f and 2f components to yield a retardation induced by the sample. Knowing the sample thickness, the retardation was recalculated into optical birefringence. Samples were prepared in 1.6-µm-thick cells with planar anchoring. The alignment quality was checked prior to measurement by inspection by polarised-light optical microscopy.

## 1.7 Dielectric studies

The complex dielectric permittivity, ε*, was measured using a Solartron 1260 impedance analyser, in the 1 Hz −10 MHz frequency range, and a probe voltage of 50 mV. The material was placed in a 10-µm-thick glass cell with gold electrodes. Cells without polymer aligning layers were used, as the presence of the thin (~10 nm) polyimide layers at the cell surfaces acts as an additional high capacitance capacitor in a series circuit with the capacitor filled with the LC sample, though this can also occur even without aligning layers, which for materials with very high values of permittivity, may strongly affect the measured permittivity of the LC phases [29]. Lack of a surfactant layer resulted in a random configuration of the director in the LC phases.

Dielectric loss data was fitted using the sum of a power law function for conductivity in the SmA phase, a Cole-Cole function [42] for the dielectric processes observed and power laws to account for any contributions out of range of the frequency window either at the short or long frequency ends of the spectrum. In practice the dielectric response in the SmA phase was fitted using:

$$\varepsilon^* = -\frac{\sigma_{DC}}{\varepsilon_O \omega} i + \frac{\delta\varepsilon_1}{(1 + (i\omega\tau_1)^{\alpha_1}} + A\omega^n i$$

where $\sigma_{DC}$ is the DC-conductivity, δε the dielectric strength parameter, τ the characteristic relaxation timescale, α the symmetric broadening parameter (value allowed to vary between 0 and 1), A the amplitude of the long high frequency powerlaw and n a fitting parameter set between 0 and 1.

In the $SmA_F$ and $SmC_P$ phases the data was fitted using:

$$\varepsilon^* = -A\omega^{-n} i + \frac{\delta\varepsilon_2}{(1 + (i\omega\tau_2)^{\alpha_2}} + \frac{\delta\varepsilon_3}{(1 + (i\omega\tau_3)^{\alpha_3}}$$

where the DC conductivity term as being replaced with a general power law to account for the mixture of conductivity and any peak contribution at low frequencies out of the frequency window. Initially, a Havrilliak-Negami function [43] was used to fit the relaxation processes however from the initial fit this was shown to be unnecessary as the β parameter of the Havrilliak-Negami function was shown to be equal to ~1 at all temperatures for all processes and so the Cole-Cole function was used instead. Representative fits are shown in Fig. S13

## 1.8 SHG Microscopy

The SHG response was investigated using a microscopic setup based on a solid-state laser EKSPLA NL202. Laser pulses (9 ns) at a 10 Hz repetition rate and max. 2 mJ pulse energy at λ= 1064 nm were applied. The pulse energy was adjusted for each sample to avoid its decomposition. The infra-red beam was incident onto a LC homogenous cell of thickness 5 µm. An IR pass filter was placed at the entrance to the sample and a green pass filter at the exit of the sample.

## 1.9 Molecular modelling

### 1.9.1 Electronic Structure Calculations

All electronic structure calculations were performed using the Gaussian G16 software package, revision c01 [44]. Geometry optimisation was performed using the B3LYP hybrid functional [45,46], the cc-pVTZ basis set [47] and GD3BJ dispersion correction [48]. Following geometry optimisation, a frequency calculation at the same level was used to confirm the geometry to be at a minimum.

### 1.9.2 Molecular Dynamics

The initial geometry for **PPZGU-3-F** (n=3) was obtained by optimisation (at the B3LYP-GD3BJ/cc-pVTZ level of DFT, as above). RESP charges were calculated at the HF/6-31G* level [49] using Gaussian G16.c01 with the route section "iop(6/33=2) iop(6/42=6) pop(chelpg,regular)". Using Antechamber (part of AmberTools24/Amber 24 software packages [50]) the Gaussian output file was converted to .mol2 format; finally, this was parsed into Gromacs readable .itp format using the ACPYPE script [51]. Simulations were performed in Gromacs 2022.2 using GPU acceleration (Nvidia V100s) *via* CUDA 12.2.1; bonded, non-bonded, PME and update tasks were offloaded to the GPU during the production run.

Initial coordinates were generated *pseudo* randomly using the Gromacs tool *gmx insert-molecules*; we construct a low-density (~ 0.1 g $cm^3$) initial configuration consisting of 1000 molecules with random positional and orientational order. To this we then perform energy minimisation *via* steepest descent, followed by equilibration in the NVE and NVT regimes (the later at T=600K). A brief (20 ns) compression simulation with an isotropic barostat ($P$ = 100

Bar) at T=600K affords a liquid-like density isotropic state, used as a starting configuration for subsequent simulations. In the case of polar simulations, a static electric field was applied across the x-axis of the simulation for 50 ns (at T=600K) to give a near saturated polar order parameter (<*P*1> > 0.95); this was used as a starting configuration for polar simulations. In all cases the production run was >300 ns total duration.

Simulations employed periodic boundary conditions in xyz. Bonds lengths were constrained to their equilibrium values with the LINCS algorithm [52]. Compressabilities in xyz dimensions were set to 4.5e-5, with the off-diagonal compressibilities were set to zero to ensure the simulation box remained rectangular. Long-range electrostatic interactions were calculated using the Particle Mesh Ewald method with a cut-off value of 1.2 nm. A van der Waals cut-off of 1.2 nm was used.

Analysis was conducted on each trajectory frame, and presented values are given as the average over all frames in a given time window (typically the entire production run), with uncertainties given as one standard deviation from the mean.

A simulation was deemed liquid crystalline if the second-rank orientational order parameter (<*P*2>), calculated *via* the Q-tensor approach using MdTraj v1.9.8 [53], was greater than 0.3 (and judged to be isotropic if below this value). Phase type was confirmed by visual inspection of the trajectory. The polar order parameter (<*P*1>) is found as the quotient of the total simulation dipole moment and the maximum possible dipole moment of the simulation. The spontaneous polarization was calculated as the ratio of the net dipole moment of the system to the simulation box volume.

In some simulations, (smectic) layer order spontaneously emerges. We measure the layer spacing and associated tilt by principal component analysis (PCA). We represent each molecule as its centre-of-mass (COM) and in doing so provide a consistent 3D coordinate for each analysis. PCA was then applied to the COM data directly, allowing the principal axes of variation in the molecular positions to be identified. By projecting the first principal component (PC1) onto a single axis, the layers are identifiable as clusters of points along said axis; thus the layer spacing is the difference between adjacent clusters. The angle between the PC1 vector (normal to the layer planes) and the director affords the tilt angle. The method detailed here has the advantage of being insensitive to the position of specific atoms and/or conformational fluctuations during the simulation.

Simulated 2D X-ray scattering patterns were computed from the simulation trajectory by modelling the electron density of each atom in the simulation as a Gaussian sphere (using atomic radii from Bondi; [54]. For all atoms in a given frame, the atomic electron density is mapped onto a 3D grid; Fourier transform of this density grid gives the 3D structure factor.

This process loops over all frames and the structure factor of the trajectory is the sum of all frames. Finally, summation of the structure factor along a given direction gives a 2D scattering pattern (e.g. summation along the [1] index corresponds to the Y-axis, thus yields the SAXS pattern in the XZ plane), following correction for the Ewald sphere. The code used for simulation of 2D SAXS patterns is available from GitHub (https://www.github.com/RichardMandle/simusaxs).

## 2 Supplementary Results

**Table S1** Transition temperatures (℃) and associated enthalpy changes (KJ mol$^{-1}$) for the **PPZGU-*n*-F** series.[ ] brackets indicate monotropic transitions

| *n* | m.p. | | $SmC_P$-$SmA_F$ | | $SmA_F$-SmA | | SmA-N | | N-I | |
|---|---|---|---|---|---|---|---|---|---|---|
| | T / ℃ | ΔH / KJ mol$^{-1}$ | T / ℃ | ΔH / KJ mol$^{-1}$ | T / ℃ | ΔH / KJ mol$^{-1}$ | T / ℃ | ΔH / KJ mol$^{-1}$ | T / ℃ | ΔH / KJ mol$^{-1}$ |
| **2** | 154.3 | 23.8 | [152.1] | 0.03 | [158.2] | 0.3 | 177.0 | 0.5 | 249.1 | 0.5 |
| **3** | 132.0 | 23.8 | [118.8] | 0.06 | 141.4 | 0.9 | 186.2 | 0.5 | 249.8 | 0.7 |
| **4** | 105.4 | 21.4 | [106.3] | 0.04 | [109.9] | 0.2 | 195.6 | 0.9 | 237.9 | 0.5 |
| **5** | 100.9 | 20.4 | - | - | - | - | 204.1 | 1.1 | 236.1 | 0.7 |
| **7** | 79.3 | 18.4 | - | - | - | - | 210.2 | 1.4 | 224.3 | 0.6 |
| **9** | 72.4 | 27.0 | - | - | - | - | 210.3 | 1.8 | 214.1 | 0.6 |

**Table S2** Calculated DFT parameters; dipole angle refers to the angle between the dipole vector and the molecular long axis.

| *n* | Dipole moment (μ) / D | Dipole Angle ($\theta_\mu$) / ° | Molecular length (L) / Å | Molecular Width (W) / Å | Molecular Aspect Ratio (Ar) |
|---|---|---|---|---|---|
| **2** | 6.95 | 11.27 | 22.28 | 5.13 | 4.34 |
| **3** | 6.99 | 10.65 | 23.26 | 5.17 | 4.50 |
| **4** | 7.09 | 11.77 | 24.57 | 5.14 | 4.78 |
| **5** | 7.07 | 11.75 | 25.58 | 5.15 | 4.96 |
| **7** | 7.10 | 13.46 | 27.95 | 5.30 | 5.27 |
| **9** | 7.12 | 15.25 | 30.36 | 5.97 | 5.07 |

The following supplemental data belongs to **PPZGU-3-F**. Data for the remaining members of the series is available at the associated DOI (quote DOI).

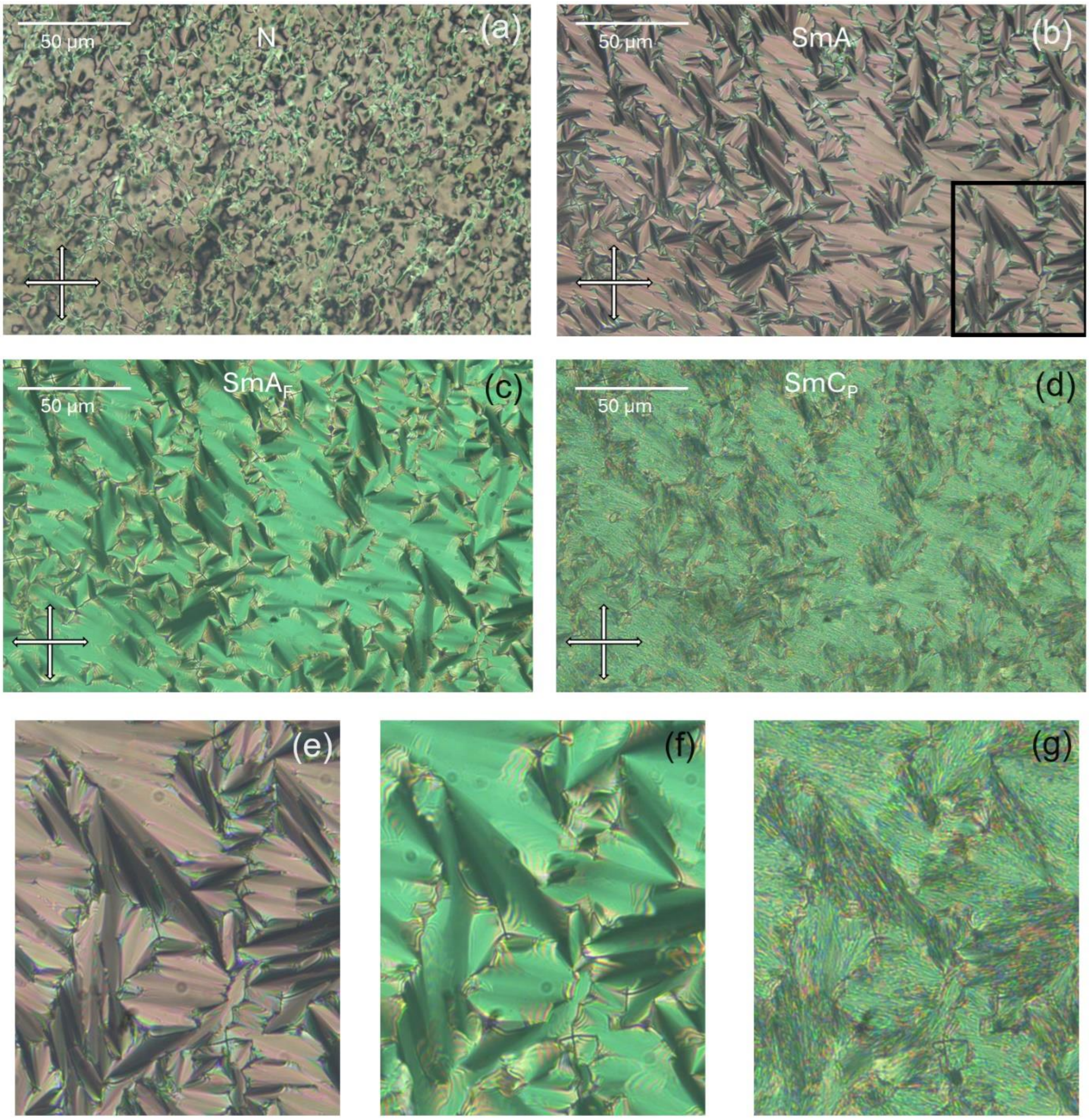

**Fig. S1.** POM micrographs of the (a) N , (b) SmA, (c) $SmA_F$ and (d) $SmC_P$ phases observed for **PPZGU-3-F** within a thin cell treated for homeotropic anchoring. The bottom row show the same black boxed region of the SmA phase in (a) for the (e) SmA, (f) $SmA_F$ and (g) $SmC_P$ phases.

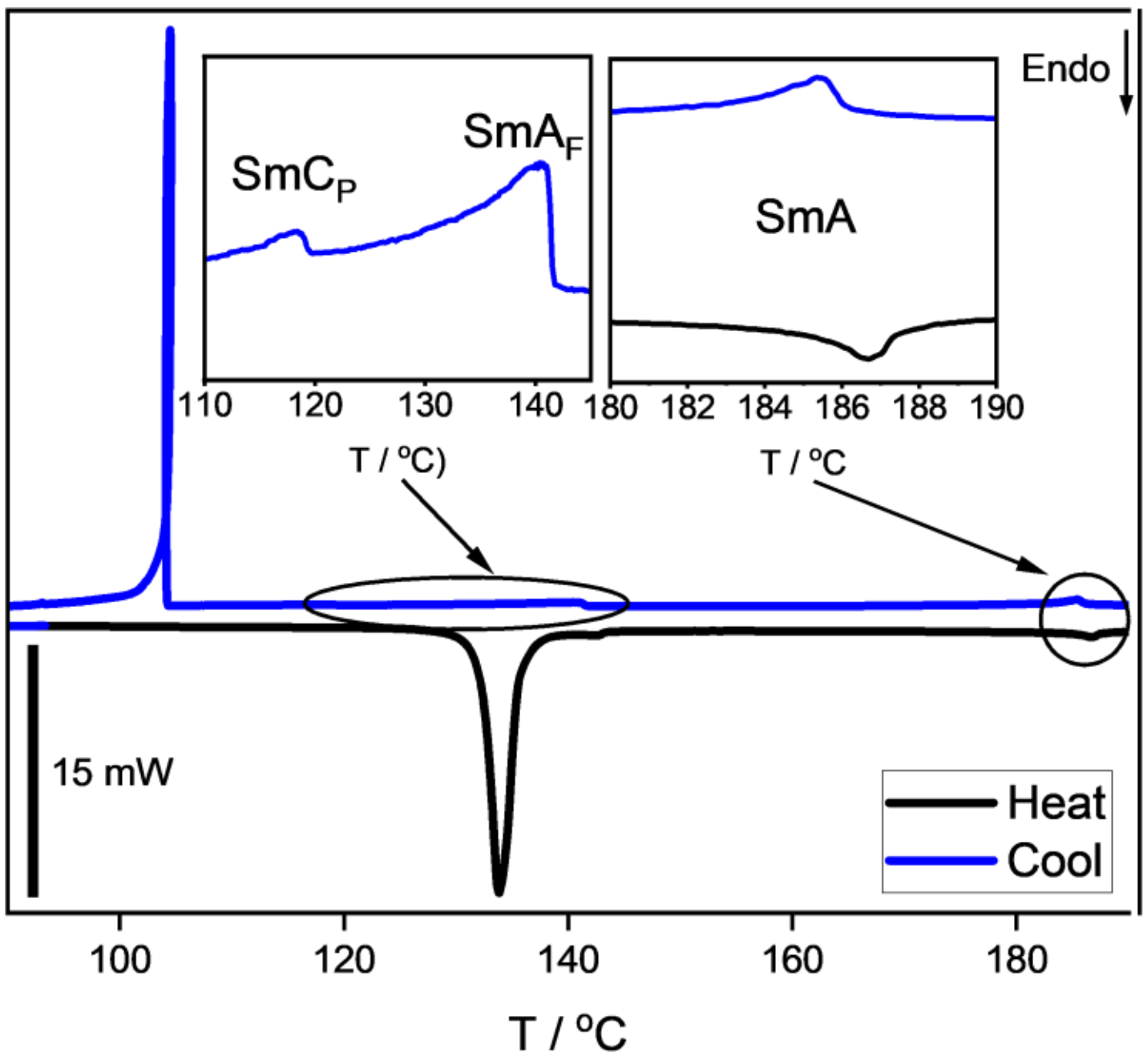

**Fig. S2.** Example DSC thermogram for **PPZGU-3-F**. Samples were measured with 10 K min$^{-1}$ cooling rate. The N-Iso transition has been omitted for clarity.

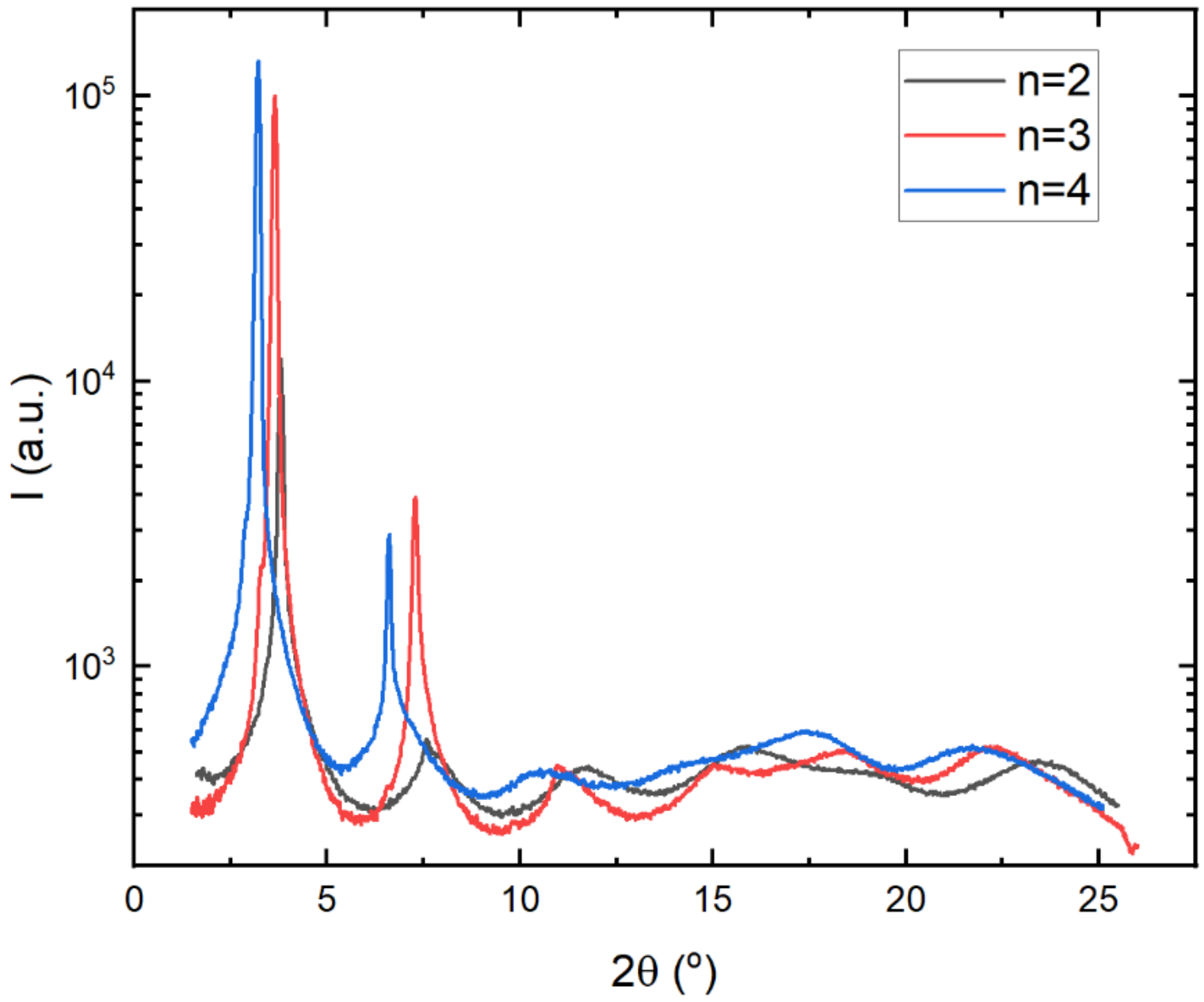

**Fig. S3.** X-ray scattering data for the polar materials showing the series of commensurate signals corresponding to layer spacing. Radial integration was done in a radial sector around layer normal direction, that exclude any wide-angle scattering.

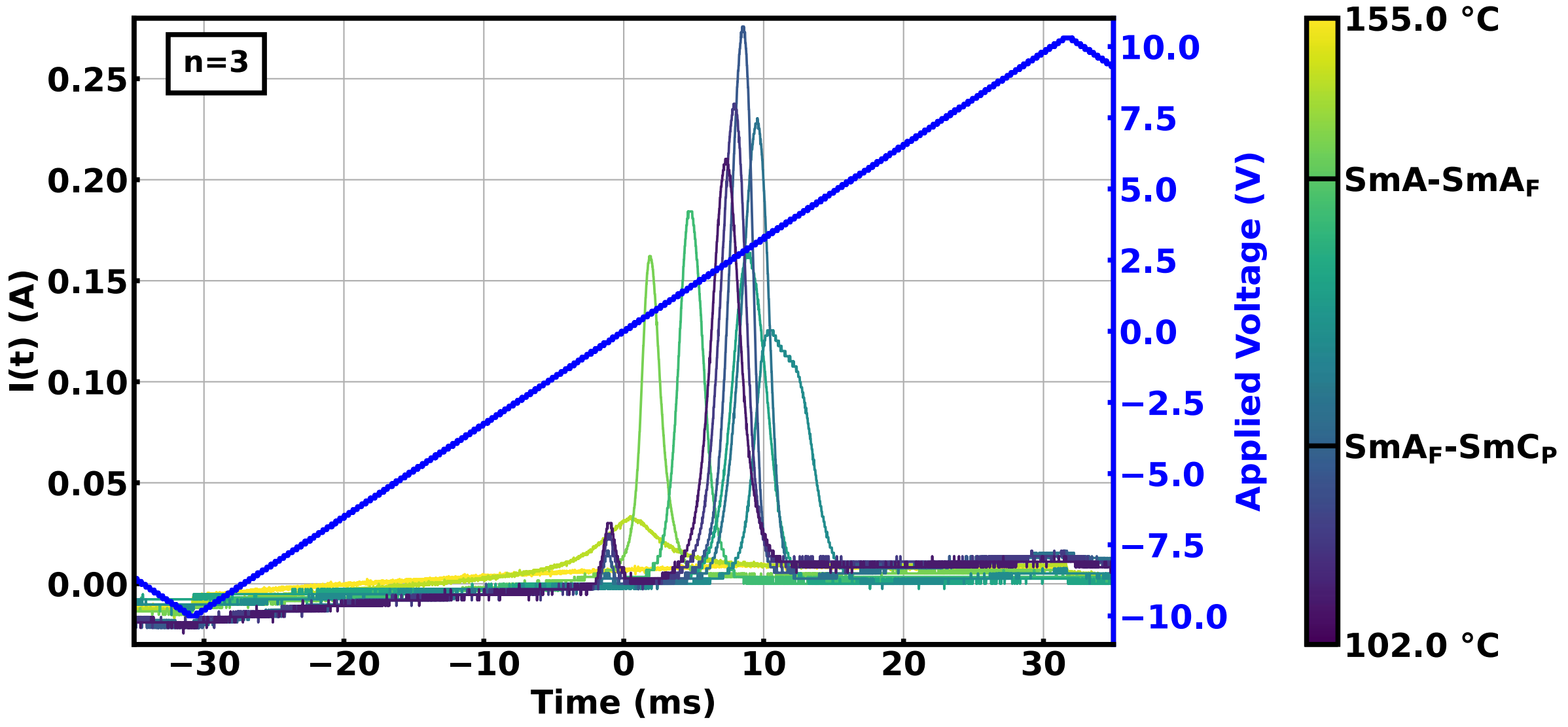

**Fig. S4.** Current response traces, measured at 8 Hz, as function of temperature for **PPZGU-3-F.**

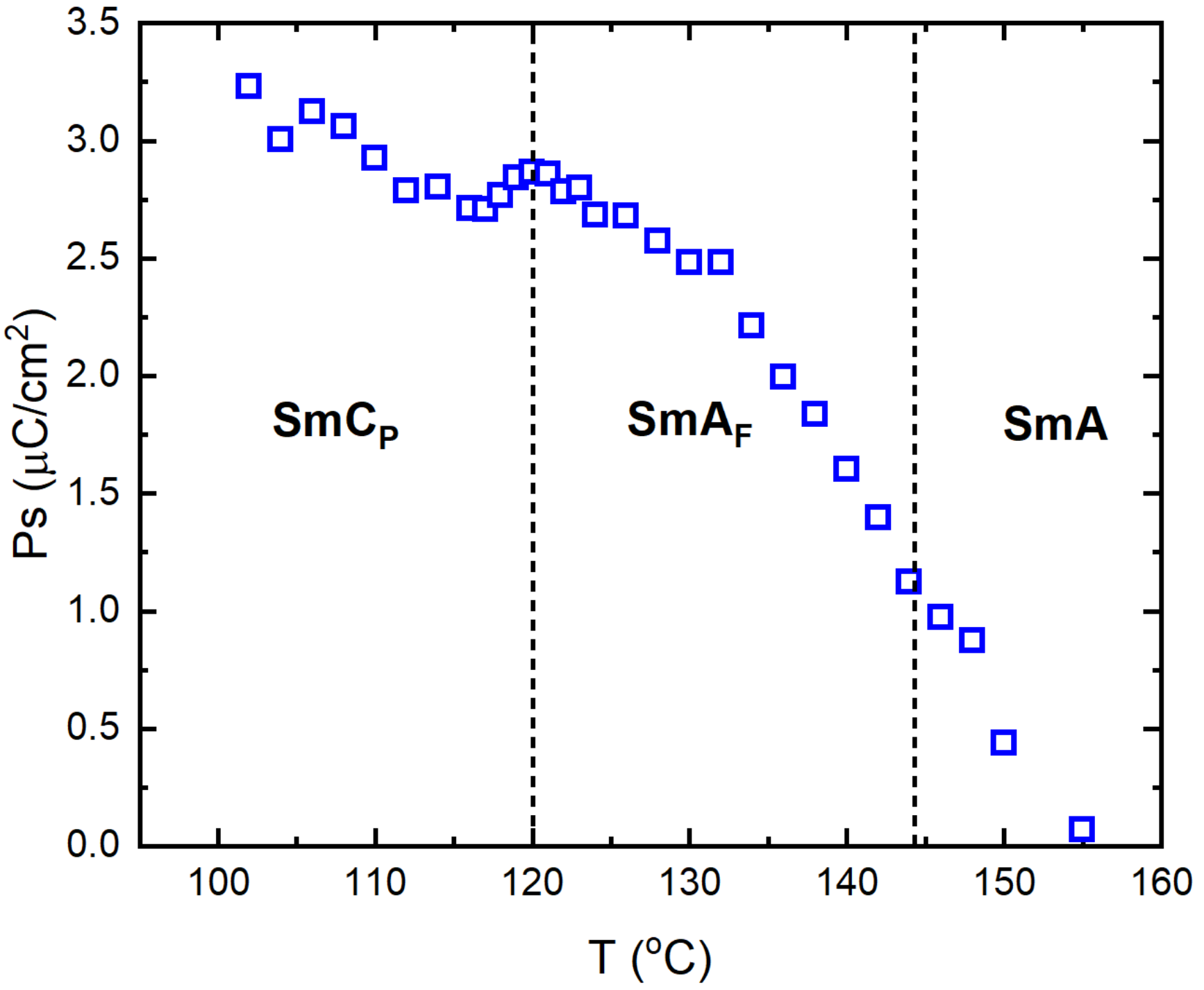

**Fig. S5** Temperature dependence of the spontaneous polarization (Ps) measured at 8 Hz in the SmA, $SmA_F$ and $SmC_P$ phases for **PPZGU-3-F**; the pre-transitional $P_S$ measured in the SmA is field-induced polarisation.

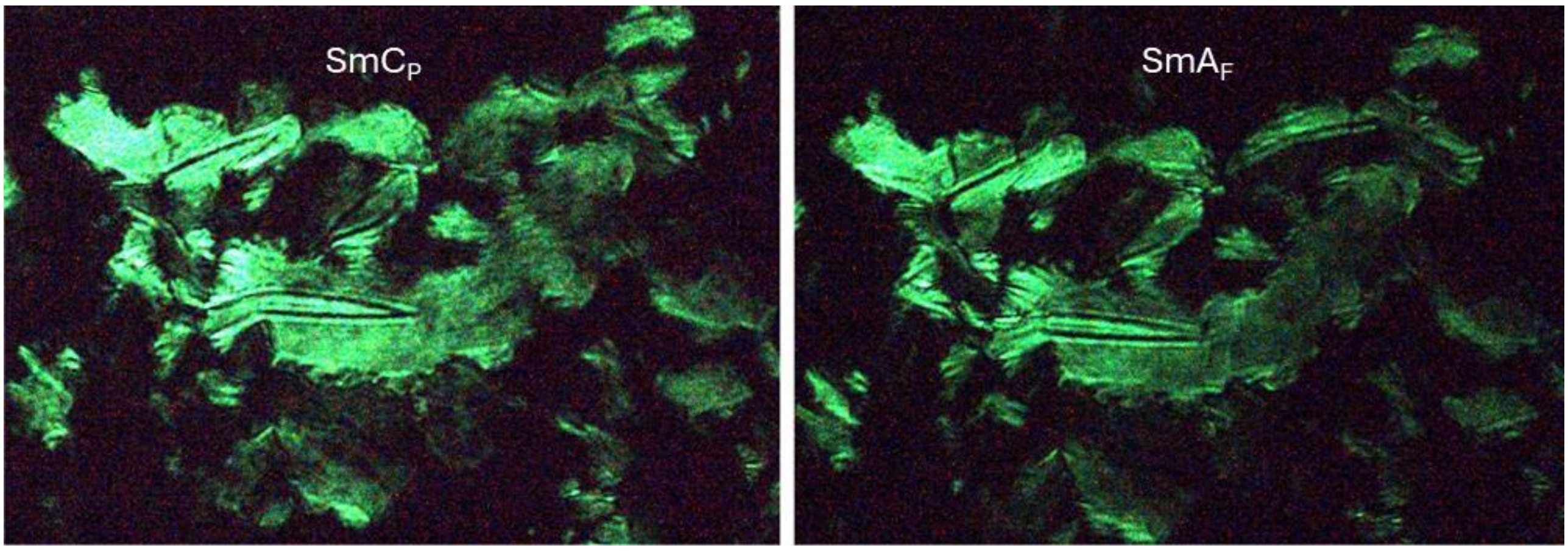

**Fig. S6.** SGH microscopy images of the SmC$_P$ [left] and SmA$_F$ [right] phases observed for **PPZGU-3-F**. The images are taken without an applied electrical field. The bright domains are those with polarization perpendicular the light propagation direction.

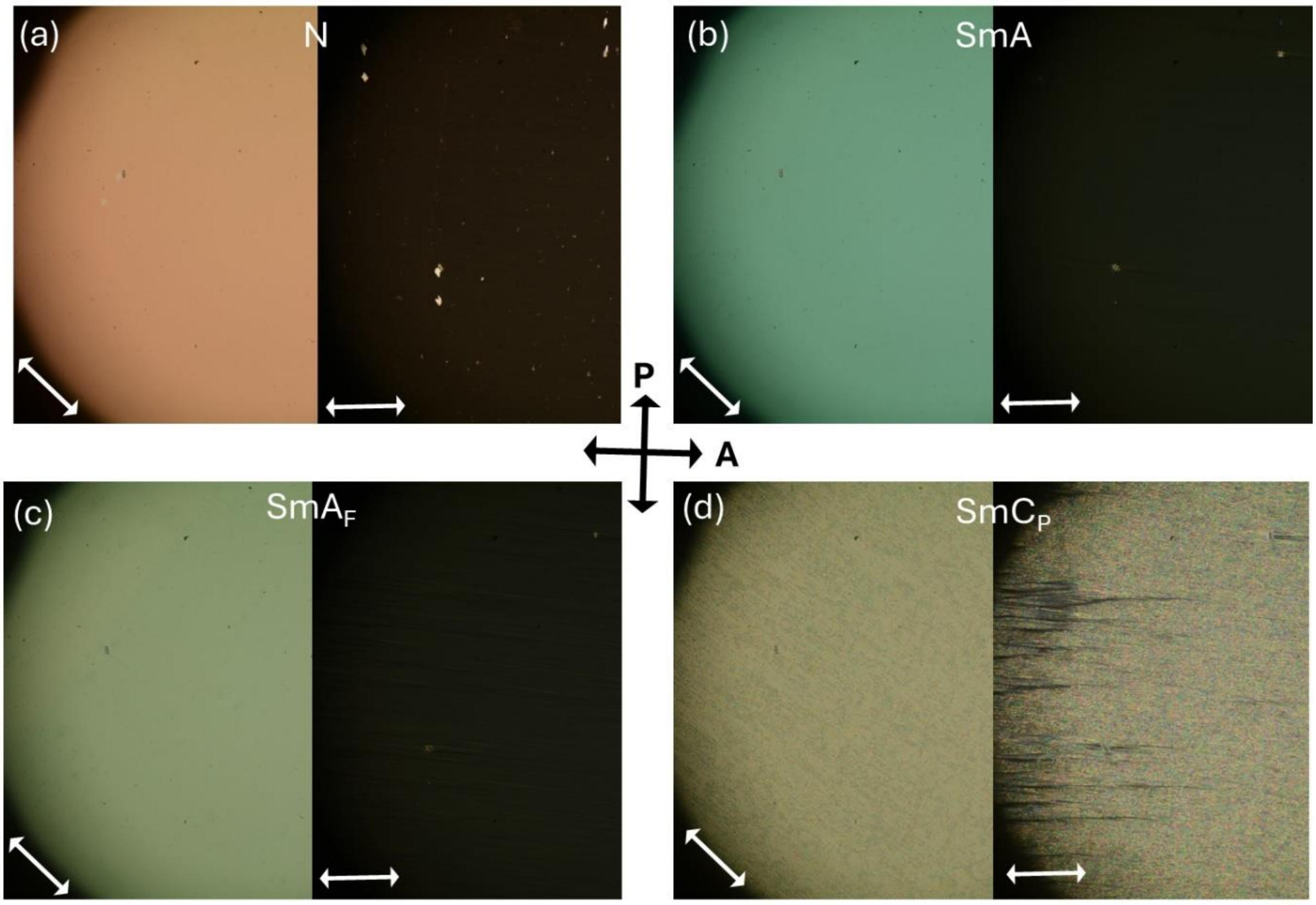

**Fig. S7.** POM micrographs comparing the light and dark states of the (a) N, (b) SmA, (c) SmA$_F$, and (d) SmC$_P$ phases. **PPZGU-3-F** samples are confined in thin cells treated for a planar anchoring condition with anti-parallel rubbing. “P” and “A” arrows indicate polariser and analyser directions respectively. White arrow on POM images indicates the rubbing direction.

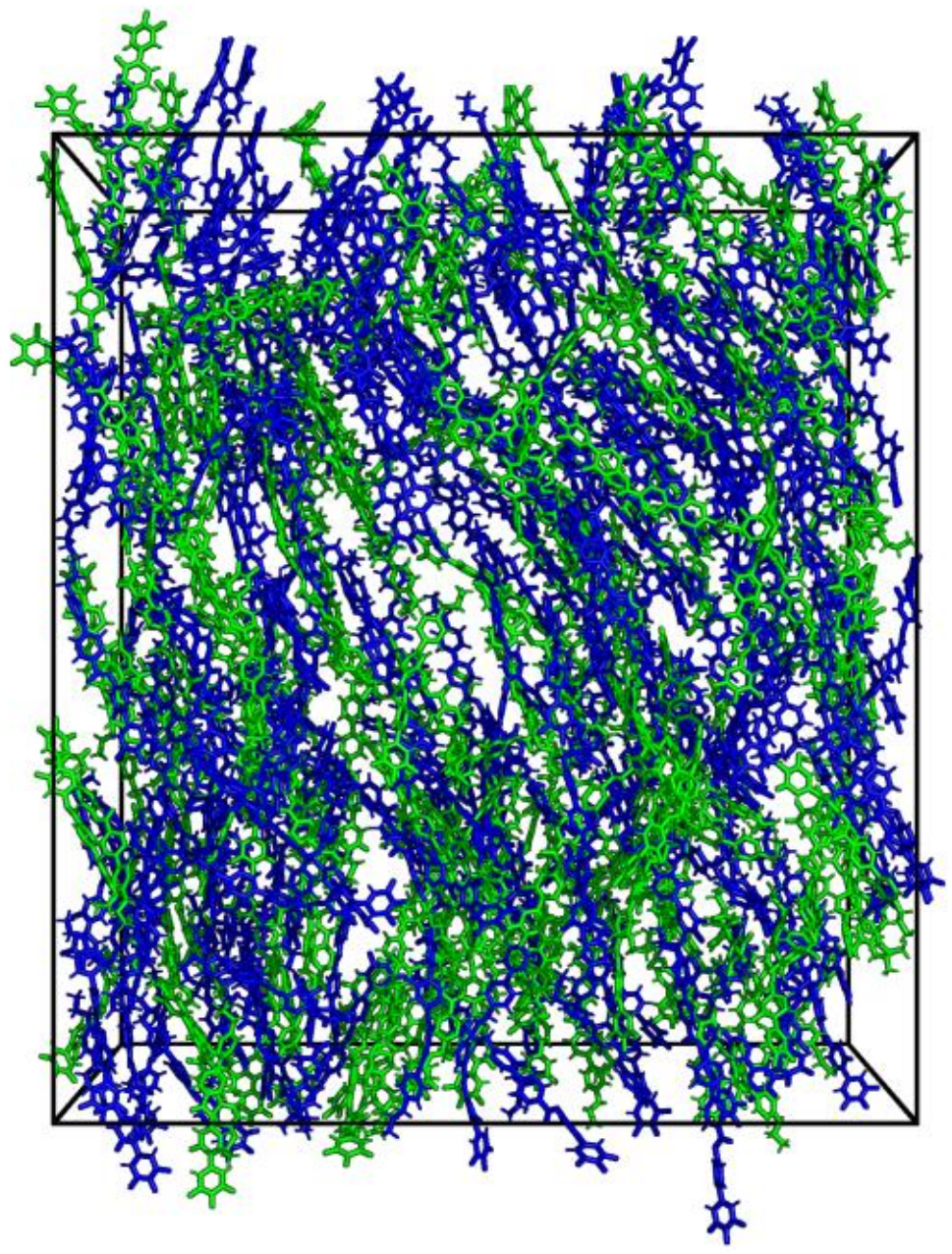

**Fig. S8.** Instantaneous configurations of: (a) the *apolar* nematic phase of **n=3** that forms spontaneously from the isotropic liquid at T 475 K at 78 ns (of a production run of >300 ns with average <*P*2> = 0.61 +/- 0.04). The snapshot is color-coded green/blue to show molecules whose dipole is parallel or antiparallel with the director, respectively (<*P*1> = 0.03 +/- 0.01). Only 400 molecules (out of 1000) are shown, to aid visualisation.

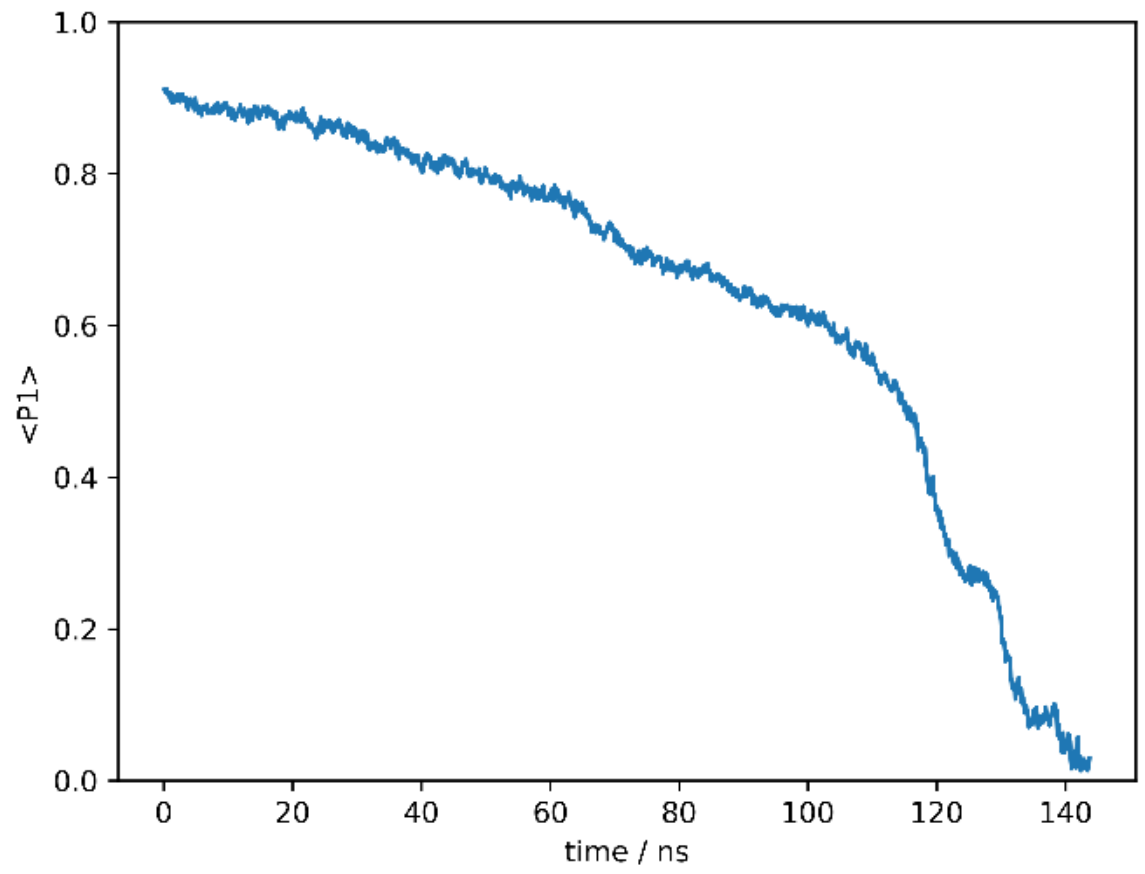

**Fig. S9.** The decay of the polar order parameter (<P1>) of PPZGU-3-F (n=3) when beginning from a field induced polar nematic configuration at T = 500 K. After removal of the field, <P1> decays to zero by t = 140 ns, at which point the simulation is in a non-polar nematic configuration (<P2> = 0.50.

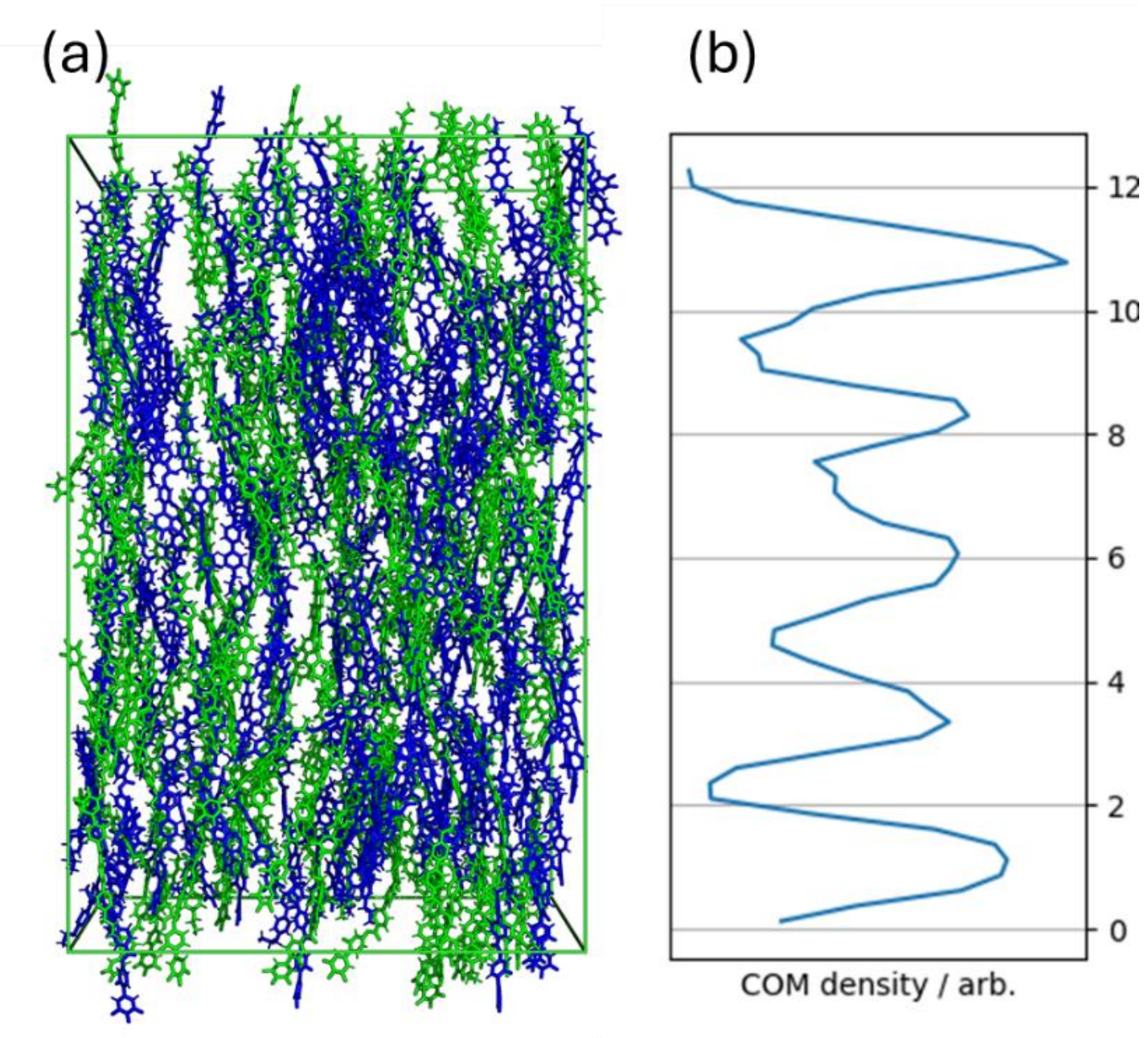

**Fig. S10.** (a) Instantaneous configuration of the *apolar* smectic A phase of PPZGU-3-F (n=3) formed spontaneously from the isotropic liquid at T 425 K at 350 ns (of a production run of >500 ns, <*P*2> = 0.81 +/- 0.08). The snapshot is color-coded green/blue to show molecules whose dipole is parallel or antiparallel with the director, respectively (<*P*1> = 0.04 +/- 0.02). The layer spacing is 2.4 +/- 0.1 nm. Only 400 molecules (out of 1000) are shown, to aid visualisation. (b) Plot of the density of molecular centres-of-mass along a vector normal to the layer plane shows the diffuse layer structure.

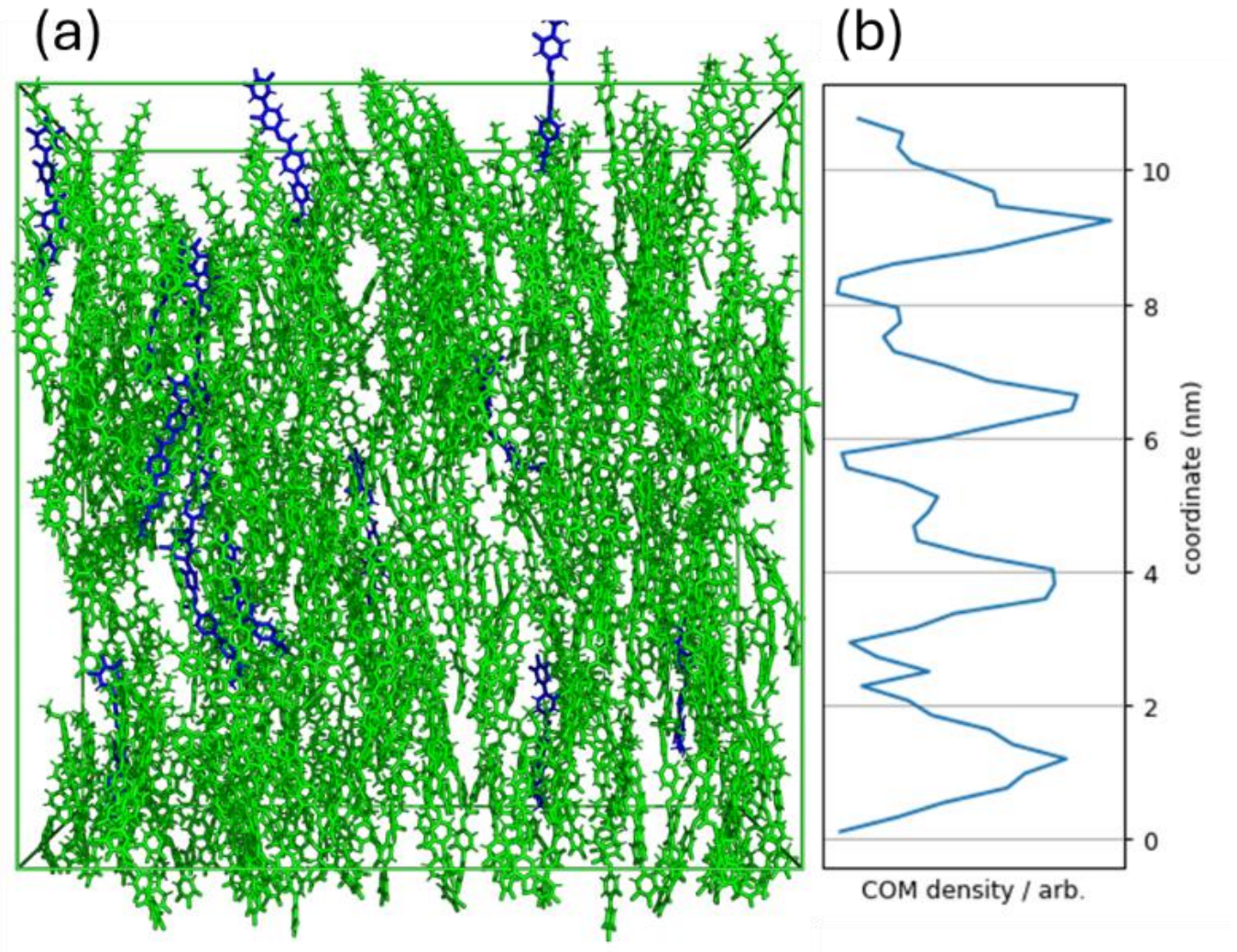

**Fig S11.** (a) Instantaneous configuration of the *polar* smectic A phase of PPZGU-3-F (n=3) formed spontaneously from the polar initial configuration at T 475 K at 259 ns (of a production run of >500 ns, <*P*2> = 0.72 +/- 0.03). The snapshot is color-coded green/blue to show molecules whose dipole is parallel or antiparallel with the director, respectively (<*P*1> = 0.88 +/- 0.03). The hydrocarbon chains are omitted from the rendering to aid visualisation of the layer structure (layer spacing = 2.3 +/- 0.1 nm; The calculated spontaneous polarisation ($P_S$) is 2.5 +/- 0.02 µC.m$^2$. Only 400 molecules (out of 1000) are shown, to aid visualisation. (b) Plot of the density of molecular centres-of-mass along the vector normal to the layer plane shows the diffuse layer structure

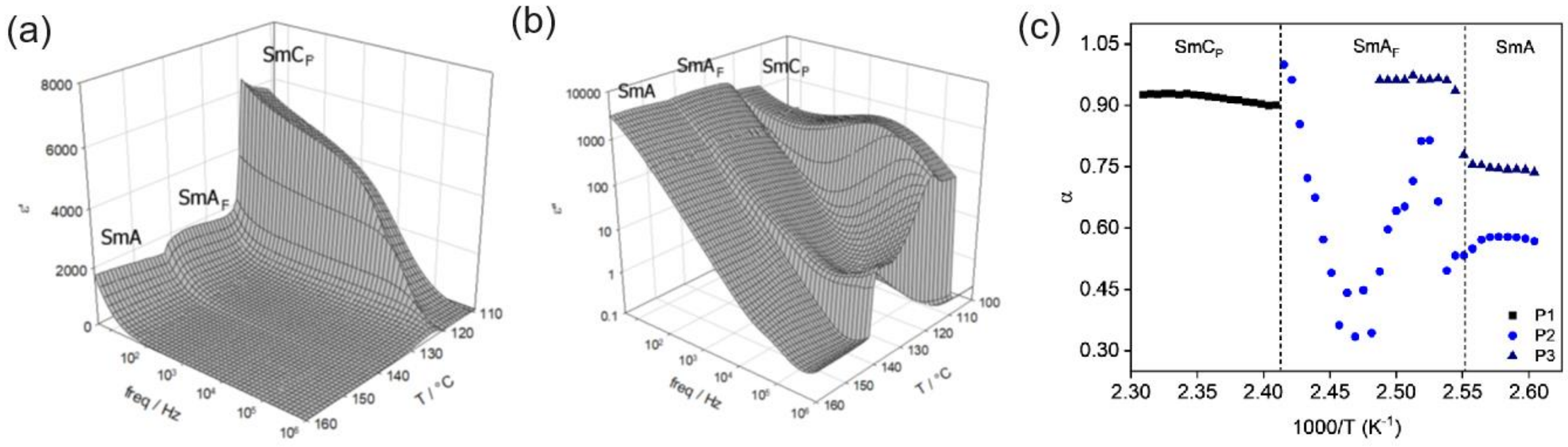

**Fig S12.** (a) Real and (b) imaginary permittivity data obtained for PPZGU-3-F in a 10 µm cell with untreated gold electrodes. (c) The α parameter obtained from fitting the dielectric loss data.

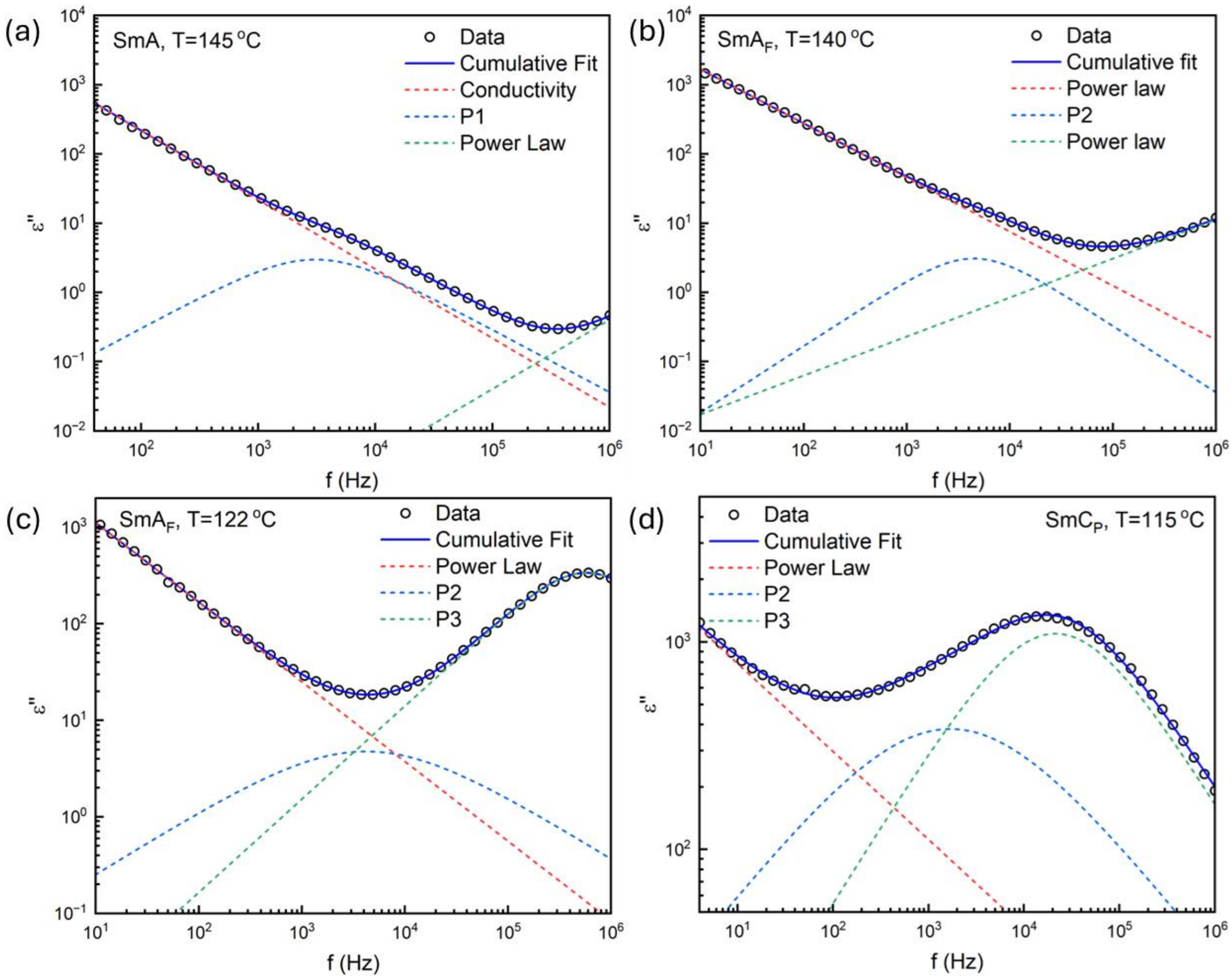

**Fig S13.** Example fits of the dielectric relaxation data obtained for PPZGU-3-F at (a) 145 °C, (b) 140 °C, (c) 122 °C and (d) 115 °C.

## 3 Organic synthesis

### 3.1 General Esterification Protocol

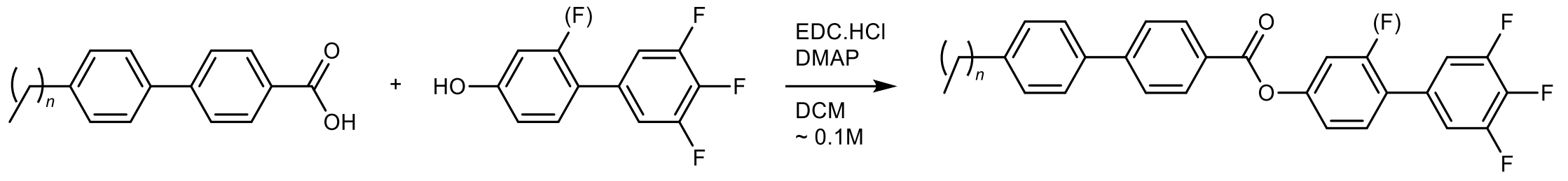

**Scheme 1.** General esterification protocol used in the preparation of the **PPZGU-*n*-F** and **PPZPU-3-F** materials.

The carboxylic acid (1.5 mmol), phenol (4-hydroxy-2,3’,4’,5’-tetrafluorobiphenyl; 1 mmol), EDC.HCl (1.5 mmol), DMAP (5 mg) were weighed into a reaction vial or round bottom flask. DCM was added (~ 10 ml, final concentration w.r.t. phenol ~ 0.1 M), and the resulting suspension stirred until complete consumption of the starting phenol as judged by TLC. The crude reaction mixture was purified by flash chromatography using a 12 g $SiO_2$ cartridge as the stationary phase and a gradient of hexane/EtOAc (0% EtOAc à 25% EtOAc) as the mobile phase, with detection made in the range 200-800 nm. The chromatographed material was filtered through a 200 nm PTFE syringe filter, concentrated to dryness, and finally recrystalised from the indicated solvent system.

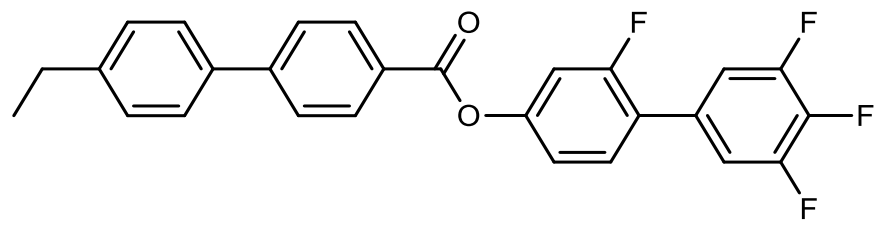

**1** (**PPZGU-2-F**): 3-fluoro-4-(3,4,5-trifluorophenyl)phenyl 4-(4-ethylphenyl)benzoate

Yield: (white crystals) 360 mg, 80%

$R_F$ (DCM): 0.76

$^1$H NMR (400 MHz, $CDCl_3$) (δ): 8.26 (ddd, J = 8.4, 1.9, 1.8 Hz, 2H, Ar-**H**), 7.75 (ddd, J = 8.5, 1.9, 1.8 Hz, 2H, Ar-**H**), 7.60 (ddd, J = 8.2, 2.2, 1.7 Hz, 2H, Ar-**H**), 7.44 (t, J = 8.6 Hz, 1H, Ar-**H**), 7.34 (ddd, J = 8.1, 2.1, 1.8 Hz, 2H, Ar-**H**), 7.24 – 7.11 (m, 4H, Ar-**H**), 2.74 (q, J = 7.6 Hz, 2H, Ar-C**H₂**-CH₃), 1.31 (t, J = 7.6 Hz, 3H, CH₂-C**H₃**).

$^{13}$C{$^1$H} NMR (101 MHz, $CDCl_3$) (δ): 164.63, 159.43 (d, J = 250.9 Hz), 151.82 (d, J = 11.1 Hz), 151.61 (ddd, J = 235.7, 9.6, 3.7 Hz), 146.77, 144.90, 139.50 (dt, J = 252.9, 16.0 Hz), 130.82, 130.60 (d, J = 4.0 Hz), 128.61, 127.28, 127.21, 127.13, 123.84 (d, J = 12.9 Hz), 118.35, 118.31, 113.69 – 112.89 (m), 110.79 (d, J = 25.7 Hz), 28.61, 15.55.

$^{19}$F NMR (376 MHz, $CDCl_3$) (δ): -114.61 (t, $J_{H-F}$ = 9.7 Hz, Ar-**F**), -134.21 (dd, $J_{F-F}$ = 20.6, $J_{H-F}$ = 8.9 Hz, Ar-**F**), -161.28 (tt, $J_{F-F}$ = 20.8, $J_{H-F}$ = 6.6 Hz, Ar-**F**).

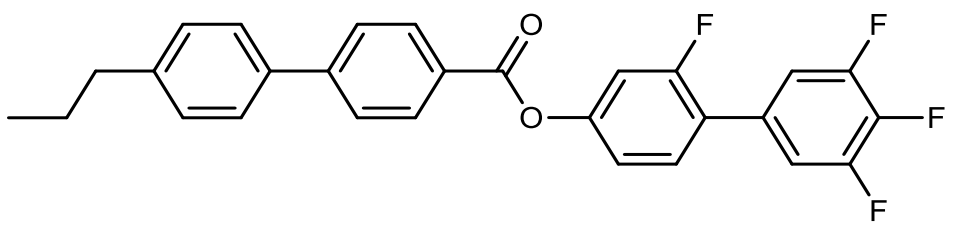

**2** (**PPZGU-3-F**): 3-fluoro-4-(3,4,5-trifluorophenyl)phenyl 4-(4-propylphenyl)benzoate

Yield: (white crystals) 357 mg, 77%

$R_F$ (DCM): 0.76

$^1$H NMR (400 MHz, $CDCl_3$) (δ): 8.25 (ddd, J = 8.7, 1.8 Hz, 2H, Ar-**H**), 7.75 (ddd, J = 8.7, 2.0, 1.8 Hz, 2H, Ar-**H**), 7.62 – 7.56 (m, 2H, Ar-**H**), 7.44 (t, J = 8.7 Hz, 1H, Ar-**H**), 7.31 (ddd, J = 8.3, 2.0, 1.9 Hz, 2H, Ar-**H**), 7.20 (ddd, J = 8.7, 6.5, 1.2 Hz, 2H, Ar-**H**), 7.16 (t, J = 2.2 Hz, 1H, Ar-**H**), 7.15 – 7.12 (m, 1H, Ar-**H**), 2.66 (t, J = 7.5 Hz, 2H, Ar-C**H₂**-CH₂), 1.70 (h, J = 7.4 Hz, 2H, CH₂-C**H₂**-CH₃), 0.99 (t, J = 7.3 Hz, 3H, CH₂-C**H₃**).

$^{13}$C{$^1$H} NMR (126 MHz, $CDCl_3$) (δ): 164.64, 160.43, 158.43, 151.82 (d, J = 11.0 Hz), 150.18 (ddd, J = 235.8, 9.7, 4.3 Hz), 146.78, 143.36, 139.50 (dt, J = 252.9, 15.2 Hz), 137.00, 130.82, 130.60 (d, J = 3.9 Hz), 129.20, 127.18, 127.12, 123.84 (d, J = 13.0 Hz), 118.35, 118.32, 113.52 – 112.87 (m), 110.79 (d, J = 25.8 Hz), 37.75, 24.54, 13.88.

$^{19}$F NMR (376 MHz, CDCl3) (δ): -114.62 (t, $J_{H-F}$ = 9.7 Hz, 1F, Ar-**F**), -134.23 (dd, $J_{F-F}$ = 20.6, $J_{H-F}$= 8.6 Hz, 2F, Ar-**F**), -161.29 (tt, $J_{F-F}$ = 20.7, $J_{H-F}$ = 6.6 Hz, 1F, Ar-**F**).

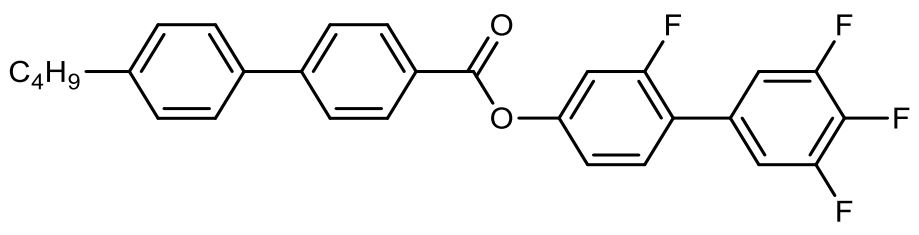

**3** (**PPZGU-4-F**): 3-fluoro-4-(3,4,5-trifluorophenyl)phenyl 4-(4-butylphenyl)benzoate

Yield: (white fluffy needles) 391 mg, 82%

$R_F$ (DCM): 0.79

$^1$H NMR (501 MHz, $CDCl_3$) (δ) : 8.26 (ddd, J = 8.2, 2.2, 1.8 Hz, 2H, Ar-**H**), 7.75 (ddd, J = 8.3, 2.0, 1.7 Hz, 2H, Ar-**H**), 7.60 (ddd, J = 8.1, 1.8, 1.7 Hz, 2H, Ar-**H**), 7.43 (t, J = 8.7 Hz, 1H, Ar-**H**), 7.33 (ddd, J = 7.9, 2.4, 2.0 Hz, 2H, Ar-**H**), 7.23 – 7.12 (m, 4H, Ar-**H**), 2.71 (t, J = 7.8 Hz, 2H, Ar-C**H₂**-CH₂), 1.68 (p, J = 7.8 Hz, 2H, CH₂-C**H₂**-CH₂), 1.44 (h, J = 7.5 Hz, 2H, CH₂-C**H₂**-CH₃), 1.00 (t, J = 7.4 Hz, 3H, CH₂-C**H₃**).

$^{13}C\{^1H\}$ NMR (126 MHz, $CDCl_3$) (δ): 164.60, 159.43 (d, J = 250.8 Hz), 151.86 (d, J = 11.0 Hz), 150.20 (ddd, J = 235.3, 9.8, 4.1 Hz), 146.75, 143.61, 139.50 (dt, J = 253.1, 14.8 Hz), 136.92, 130.82, 130.57 (d, J = 3.8 Hz), 129.17, 127.19, 127.09, 123.78 (d, J = 12.9 Hz), 118.36, 118.33, 113.51 – 112.89 (m), 110.79 (d, J = 25.8 Hz), 35.40, 33.62, 22.46, 14.00.

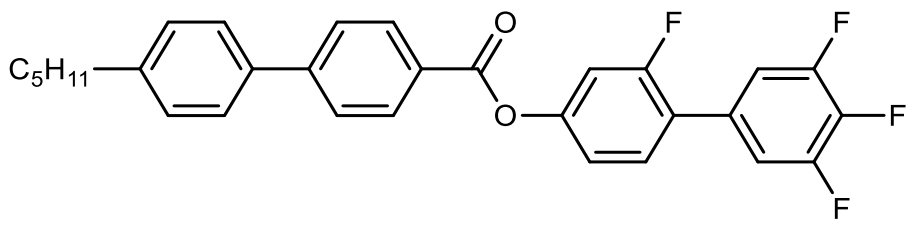

**4** (**PPZGU-5-F**): 3-fluoro-4-(3,4,5-trifluorophenyl)phenyl 4-(4-pentylphenyl)benzoate

Yield: (white crystals) 364 mg, 74%

$R_F$ (DCM): 0.80

$^1H$ NMR (400 MHz, $CDCl_3$) (δ): 8.25 (ddd, J = 8.4, 1.9, 1.8 Hz, 2H, Ar-**H**), 7.75 (ddd, J = 8.5, 2.0, 1.9 Hz, 2H, Ar-**H**), 7.59 (ddd, J = 8.0, 1.9, 1.7 Hz, 2H, Ar-**H**), 7.44 (t, J = 8.7 Hz, 1H, Ar-**H**), 7.31 (ddd, J = 8.1, 2.2, 1.8 Hz, 2H, Ar-**H**), 7.23 – 7.12 (m, 3H, Ar-**H**), 2.68 (t, J = 7.3 Hz, 2H, Ar-C**H₂**-CH₂), 1.67 (p, J = 7.5 Hz, 2H, $CH_2$-C**H₂**-$CH_2$), 1.44 – 1.30 (m, 4H, $CH_2$-(C**H₂**)₂-$CH_2$), 0.92 (t, J = 6.7 Hz, 3H, $CH_2$-C**H₃**).

$^{13}C\{^1H\}$ NMR (101 MHz, $CDCl_3$) (δ): 164.64, 159.43 (d, J = 251.0 Hz), 151.82 (d, J = 11.1 Hz), 151.22 (ddd, J = 249.7, 5.6, 3.9 Hz), 146.78, 143.63, 139.50 (dt, J = 253.1, 14.8 Hz), 136.96, 130.81, 130.60 (d, J = 4.0 Hz), 129.14, 127.19, 127.12, 123.78 (d, J = 12.5 Hz), 118.35, 118.31, 113.41 – 113.01 (m), 110.79 (d, J = 25.7 Hz), 35.65, 31.54, 31.14, 22.57, 14.05.

$^{19}F$ NMR (376 MHz, $CDCl_3$) (δ): -114.62 (t, $J_{H-F}$ = 10.0 Hz, Ar-**F**), -134.22 (dd, $J_{F-F}$ = 20.6, $J_{H-F}$ 8.6 Hz, Ar-**F**), -161.29 (tt, $J_{F-F}$ = 20.6, $J_{H-F}$ = 6.6 Hz, Ar-**F**).

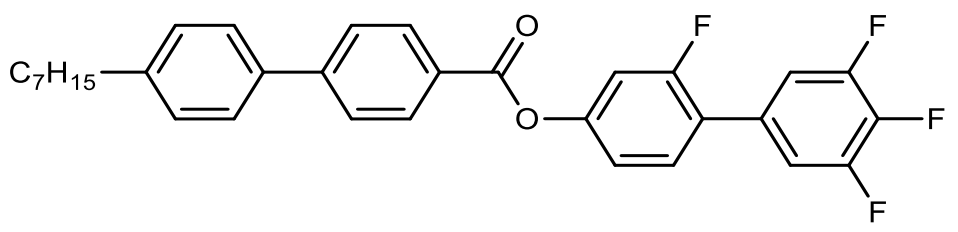

**5** (**PPZGU-7-F**): 3-fluoro-4-(3,4,5-trifluorophenyl)phenyl 4-(4-heptylphenyl)benzoate

Yield: (white solid) 406 mg, 78%

$R_F$ (DCM): 0.80

$^1$H NMR (400 MHz, $CDCl_3$) (δ): 8.25 (ddd, J = 8.5, 1.9, 1.8 Hz, 2H, Ar-**H**), 7.75 (ddd, J = 8.5, 1.9, 1.8 Hz, 2H, Ar-**H**), 7.59 (ddd, J = 8.0, 2.0, 1.7 Hz, 2H, Ar-**H**), 7.44 (t, J = 8.7 Hz, 1H, Ar-**H**), 7.31 (ddd, J = 8.2, 1.8, 1.8 Hz, 2H, Ar-**H**), 7.24 – 7.11 (m, 4H, Ar-**H**), 2.68 (t, J = 7.5 Hz, 2H, Ar-C**H₂**-CH₂), 1.67 (p, J = 7.7 Hz, 2H, CH₂-C**H₂**-CH₂), 1.45 – 1.24 (m, 8H, CH2-(C**H₂**)₄-CH₂), 0.90 (t, J = 6.8 Hz, 3H, CH₂-C**H₃**).

$^{13}$C{$^1$H} NMR(101 MHz, $CDCl_3$) (δ): 164.64, 159.43 (d, J = 250.9 Hz), 151.82 (d, J = 11.0 Hz), 149.98 (ddd, J = 243.9, 6.4, 3.9 Hz), 146.78, 143.64, 138.24 (dt, J = 253.7, 15.4 Hz), 136.94, 130.59 (d, J = 4.0 Hz), 130.57, 129.14, 127.19, 127.11, 123.84 (d, J = 13.6 Hz), 118.35, 118.31, 113.56 – 112.93 (m), 110.79 (d, J = 25.8 Hz), 35.69, 31.85, 31.48, 29.33, 29.21, 22.70, 14.13.

$^{19}$F NMR (376 MHz, CDCl3) (δ): -114.61 (t, $J_{H-F}$ = 9.8 Hz, Ar-**F**), -134.21 (dd, $J_{F-F}$ = 20.7, $J_{H-F}$ = 8.6 Hz, Ar-**F**), -161.28 (tt, $J_{F-F}$ = 20.6, $J_{H-F}$ = 6.5 Hz, Ar-**F**).

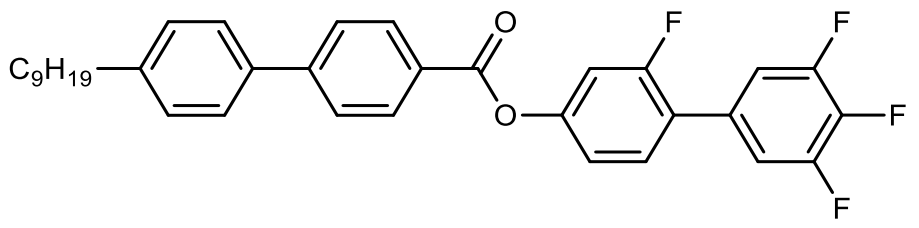

**6** (**PPZGU-9-F**): 3-fluoro-4-(3,4,5-trifluorophenyl)phenyl 4-(4-nonylphenyl)benzoate

Yield: (white solid) 384 mg, 70%

$R_F$ (DCM): 0.81

$^1$H NMR (400 MHz, $CDCl_3$) (δ): 8.26 (d, J = 8.2 Hz, 2H, Ar-**H**), 7.75 (d, J = 8.1 Hz, 2H, Ar-**H**), 7.59 (d, J = 7.8 Hz, 2H, Ar-**H**), 7.44 (t, J = 8.7 Hz, 1H, Ar-**H**), 7.32 (d, J = 7.8 Hz, 2H, Ar-**H**), 7.24 – 7.12 (m, 4H, Ar-**H**), 2.69 (t, J = 7.7 Hz, 2H, Ar-C**H₂**-CH₂), 1.68 (p, J = 7.4 Hz, 2H, CH₂-C**H₂**-CH₂), 1.46 – 1.21 (m, 14H, CH₂-(C**H₂**)₇-CH₂), 0.91 (t, J = 6.6 Hz, 3H, CH₂-C**H₃**).

$^{13}$C{$^1$H} NMR (101 MHz, $CDCl_3$) (δ): 164.62, 159.43 (d, J = 251.0 Hz), 151.84 (d, J = 11.0 Hz), 149.96 (ddd, J = 235.9, 10.0, 4.2 Hz), 146.78, 143.65, 138.24 (dt, J = 252.8, 15.2 Hz), 136.94, 130.81, 130.59 (d, J = 4.0 Hz), 129.14, 127.19, 127.11, 123.82 (d, J = 13.0 Hz), 118.35, 118.31, 113.52 – 112.85 (m), 110.79 (d, J = 25.8 Hz), 35.70, 31.94, 31.48, 29.60, 29.57, 29.38*, 22.72, 14.15.

*overlapping aliphatic signal, corresponds to two aliphatic carbon atoms.

$^{19}$F NMR (376 MHz, $CDCl_3$) (δ): -114.59 (t, $J_{H-F}$ = 9.8 Hz, Ar-**F**), -134.18 (dd, $J_{F-F}$ = 20.6, $J_{H-F}$ = 8.9 Hz, Ar-**F**), -161.25 (tt, $J_{F-F}$ = 20.9, $J_{H-F}$ = 6.6 Hz, Ar-**F**).

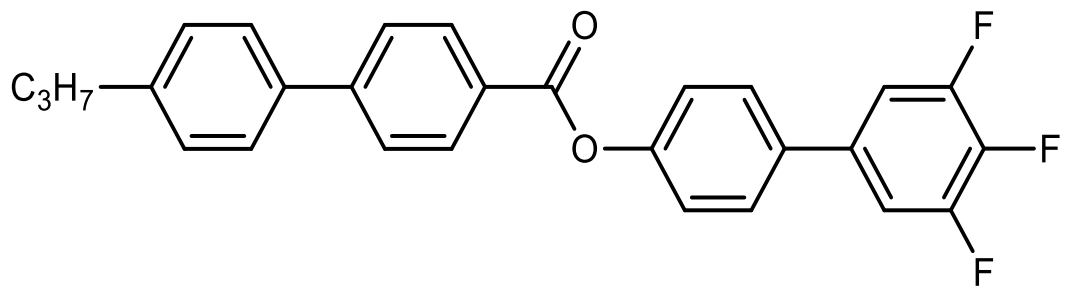

**7 (PPZPU-3-F)**: 4-(3,4,5-trifluorophenyl)phenyl 4-(4-propylphenyl)benzoate

Yield: (white solid) 336 mg, 75%

$R_F$ (DCM): 0.77

$^1$H NMR (501 MHz, $CDCl_3$) (δ): 8.27 (ddd, J = 8.4, 1.8, 1.7 Hz, 2H, Ar-**H**), 7.75 (ddd, J = 8.4, 1.9, 1.7 Hz, 2H, Ar-**H**), 7.60 (ddd, J = 8.4, 1.9, 1.7 Hz, 1H, Ar-**H**), 7.56 (ddd, J = 8.7, 2.8, 2.0 Hz, 2H, Ar-**H**), 7.36 – 7.29 (m, 4H, Ar-**H**), 7.24 – 7.16 (m, 2H, Ar-**H**), 2.67 (t, J = 6.8 Hz, 2H, Ar-C$\mathbf{H_2}$-$CH_2$), 1.71 (h, J = 7.5 Hz, 2H, $CH_2$-C$\mathbf{H_2}$-$CH_3$), 1.00 (t, J = 7.3 Hz, 3H, $CH_2$-C$\mathbf{H_3}$).

$^{13}$C{$^1$H} NMR (126 MHz, $CDCl_3$) (δ): 165.05, 152.22 (ddd, J = 235.4, 5.6, 4.2 Hz), 152.41, 146.54, 143.28, 139.34 (td, J = 252.1, 15.5 Hz), 137.09, 136.81 – 136.29 (m), 135.95, 132.55, 130.77, 129.19, 128.05, 127.62, 127.18, 127.07, 122.50, 111.12 (dd, J = 11.2, 5.4 Hz), 37.75, 24.55, 13.89.

## 3.2 Example NMR Spectra

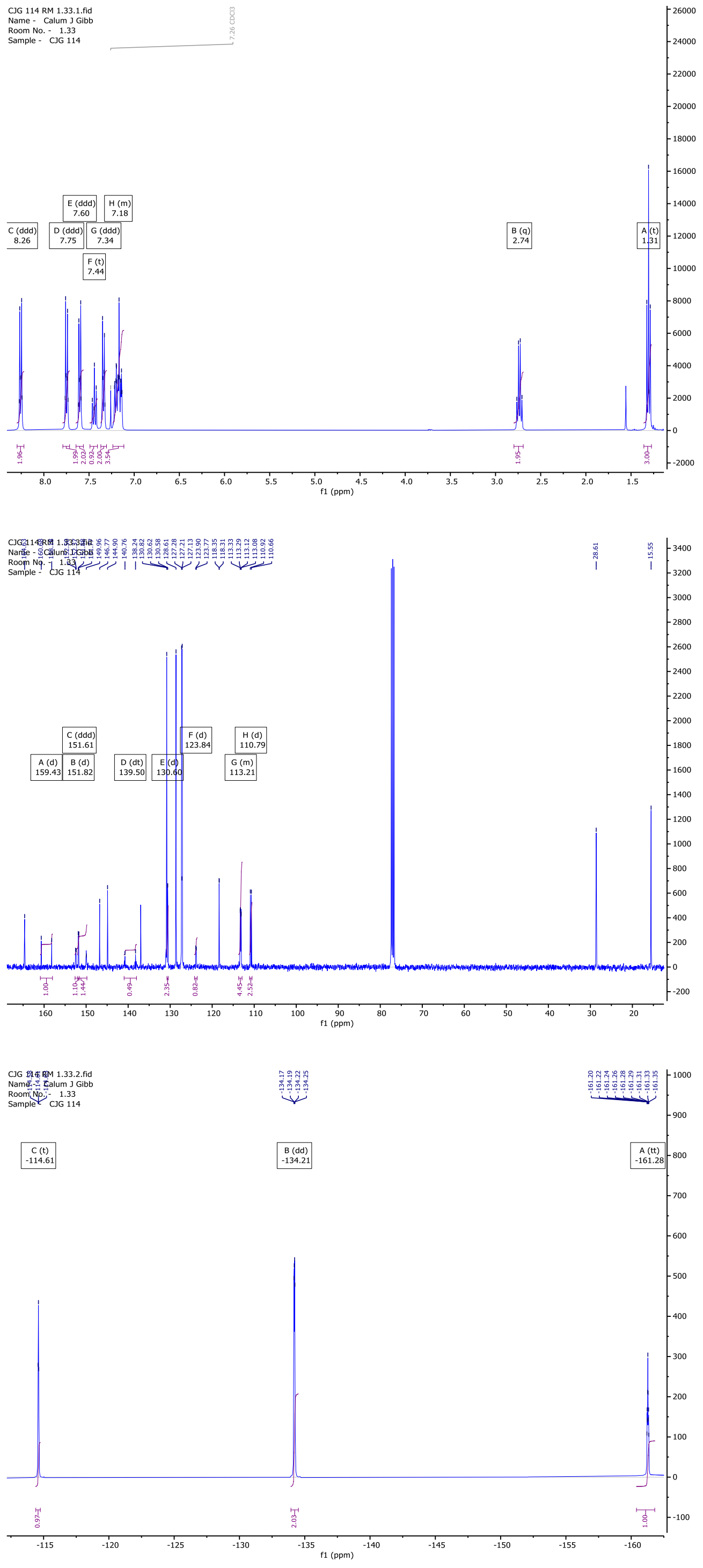

**Fig. S14.** $^{1}H$ [top], $^{13}C\{^{1}H\}$ [middle], and $^{19}F$ NMR spectra for **1** (**PPZGU-2-F**).

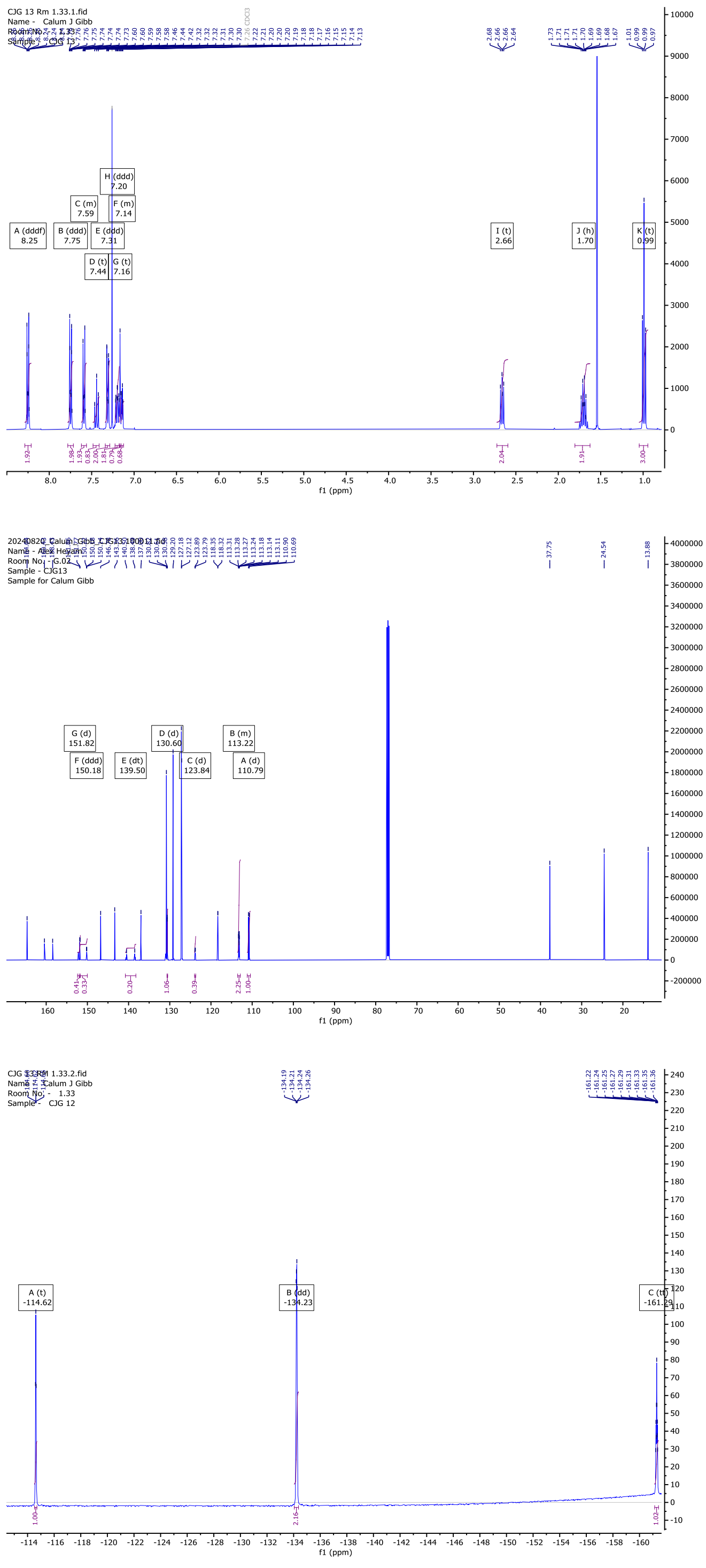

**Fig. S15.** $^{1}H$ [top], $^{13}C\{^{1}H\}$ [middle], and $^{19}F$ NMR spectra for **2** (**PPZGU-3-F**).

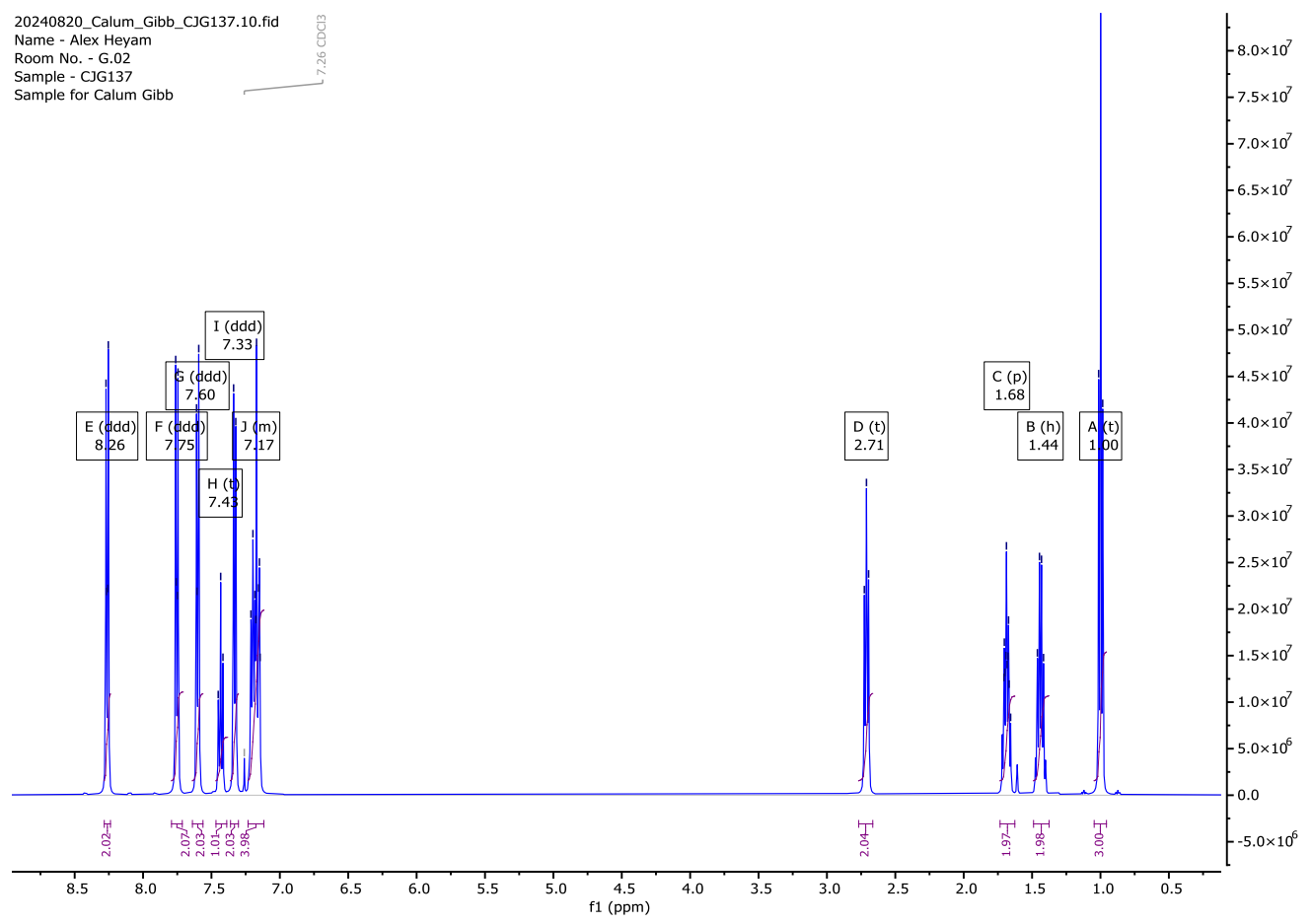

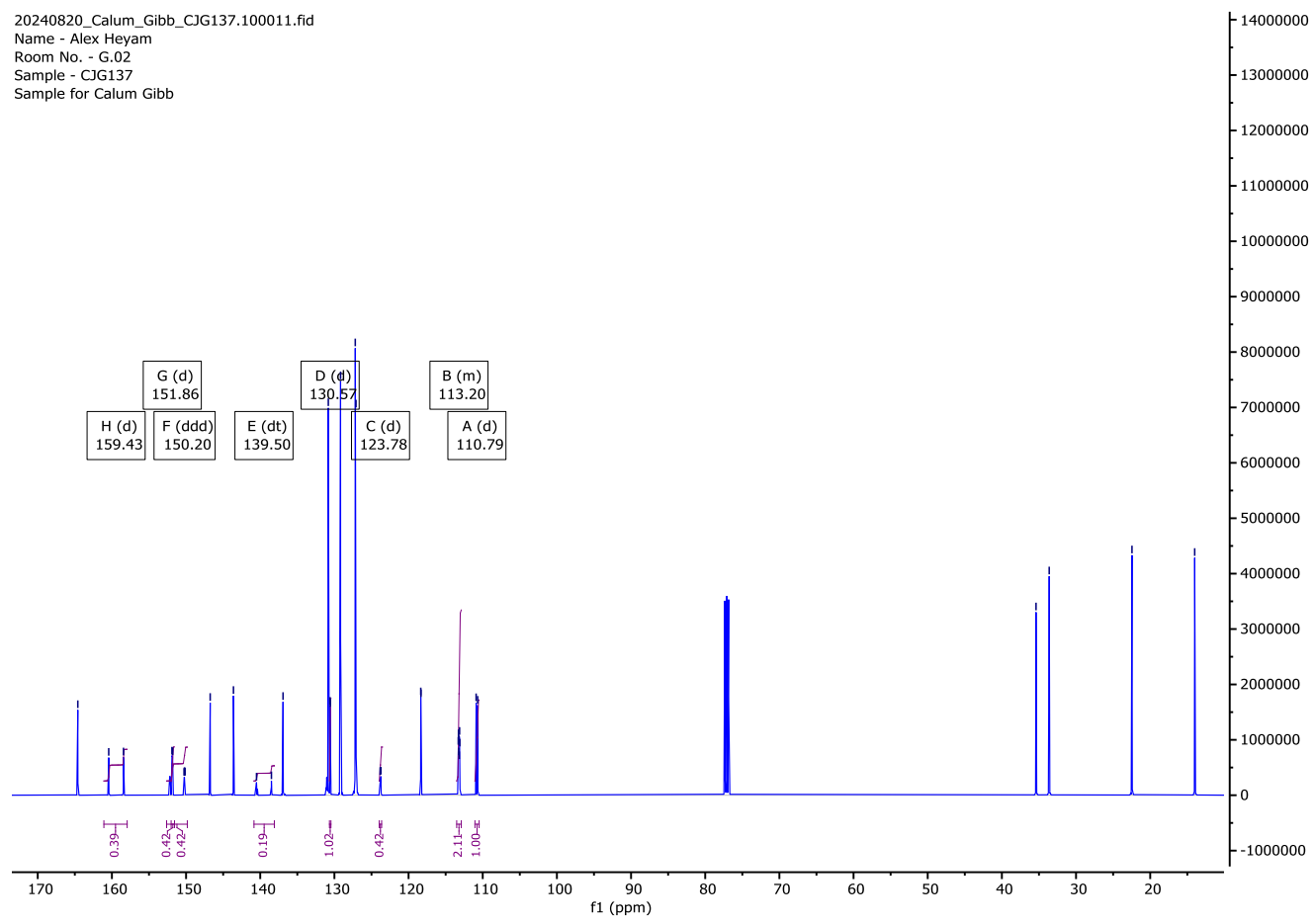

**Fig. S16.** $^{1}H$ [top], $^{13}C\{^{1}H\}$ [middle], and $^{19}F$ NMR spectra for **3** (**PPZGU-4-F**).

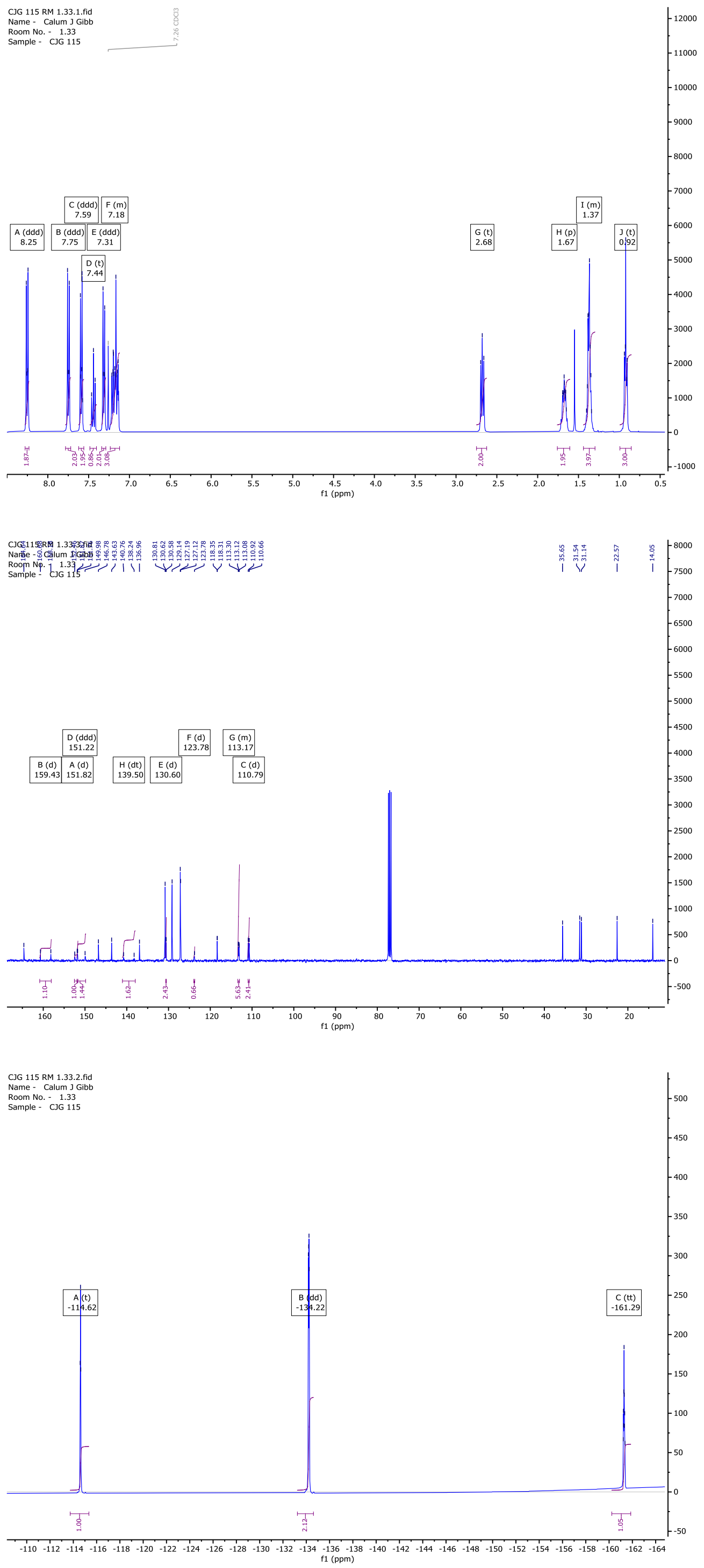

**Fig. S17.** $^{1}H$ [top], $^{13}C\{^{1}H\}$ [middle], and $^{19}F$ NMR spectra for **4** (**PPZGU-5-F**).

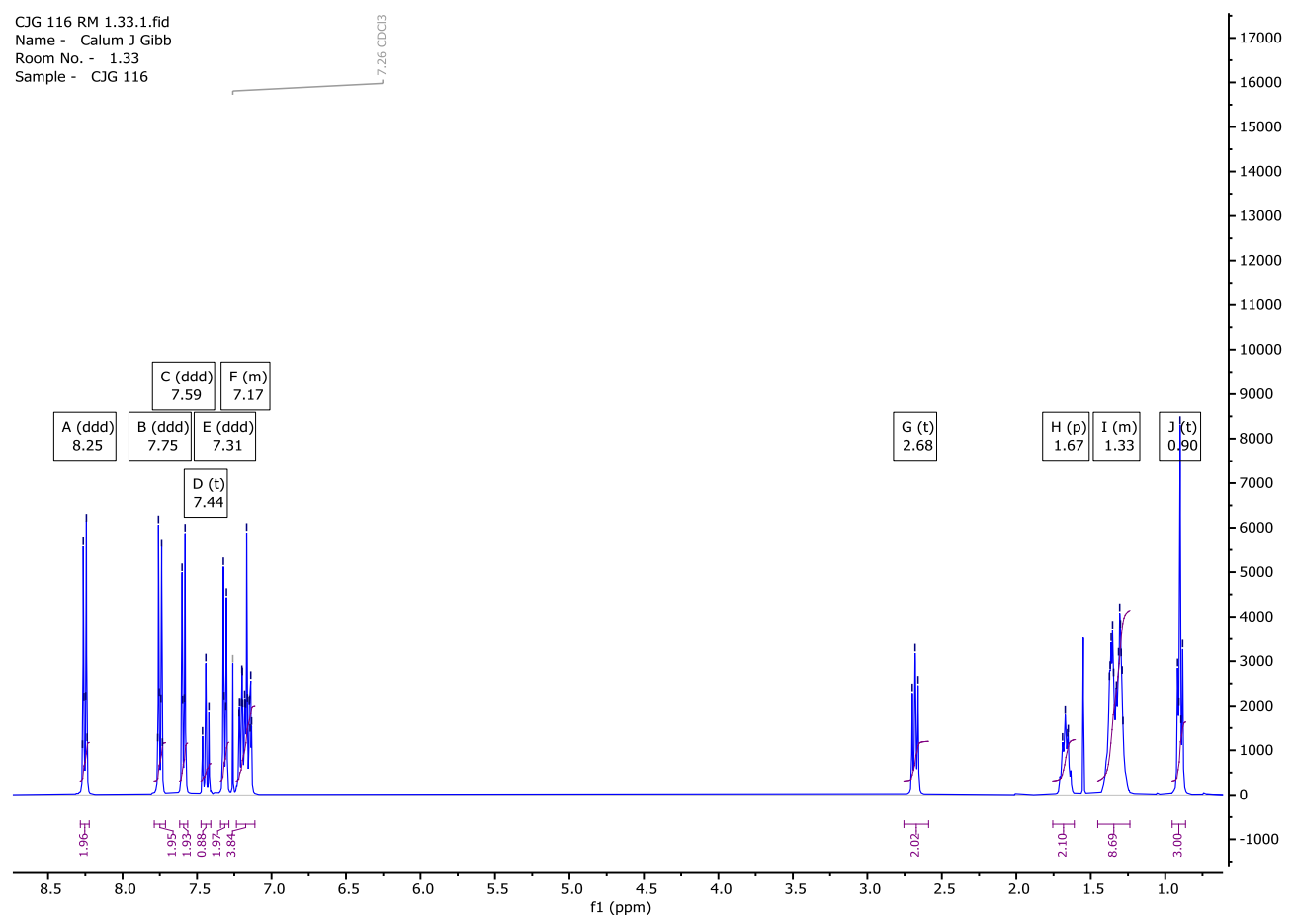

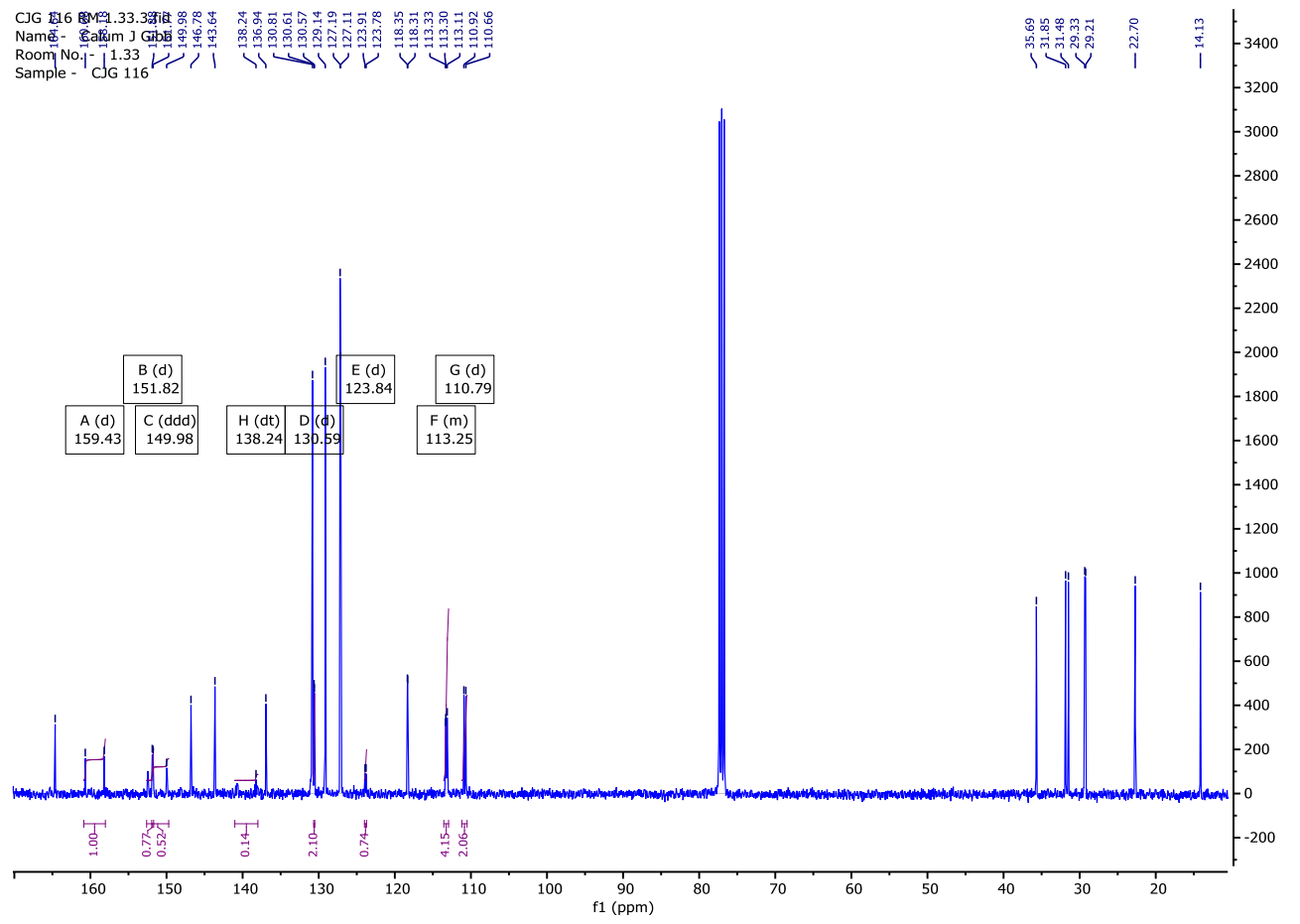

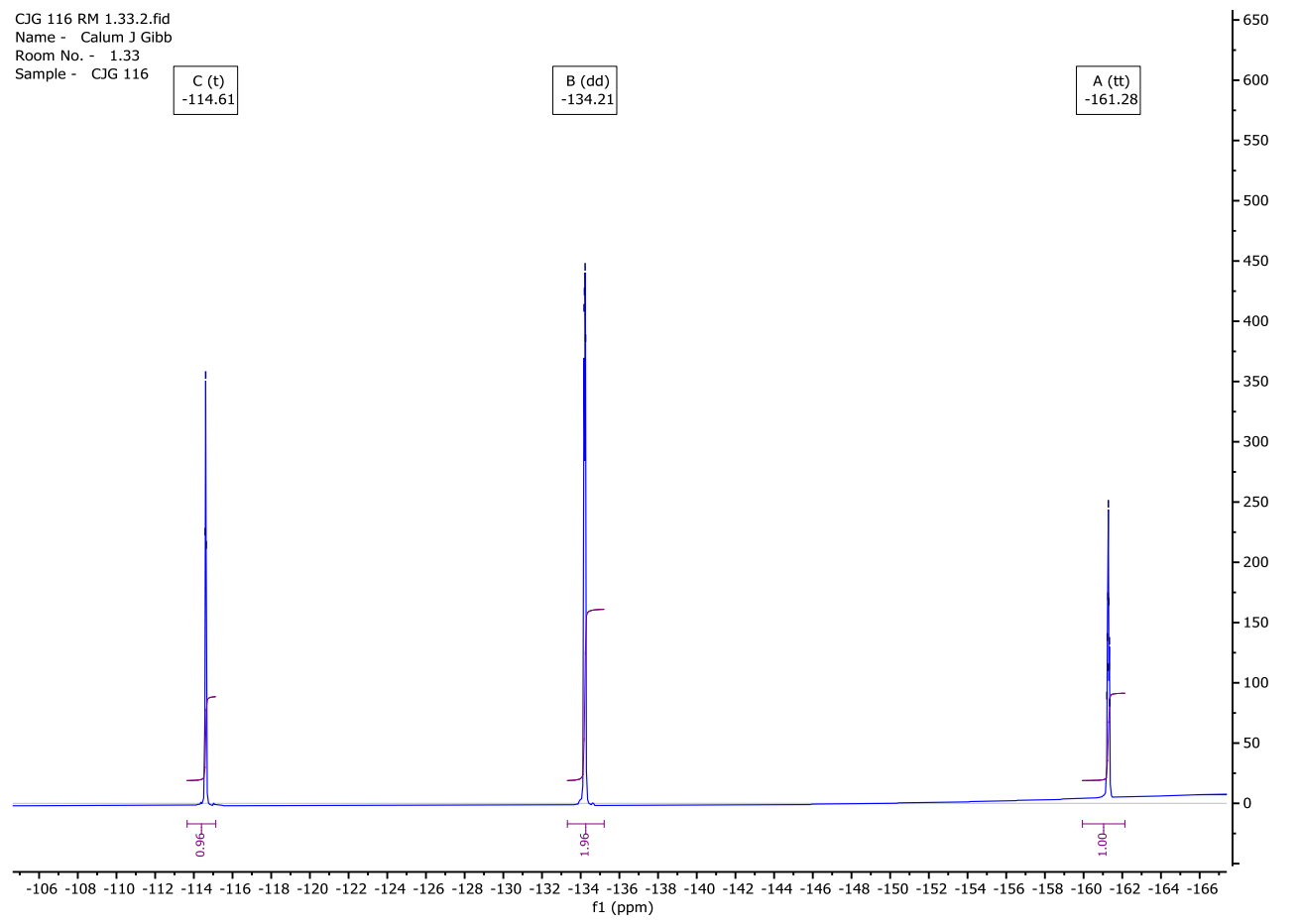

**Fig. S18.** $^{1}$H [top], $^{13}$C{$^{1}$H} [middle], and $^{19}$F NMR spectra for **5** (**PPZGU-7-F**).

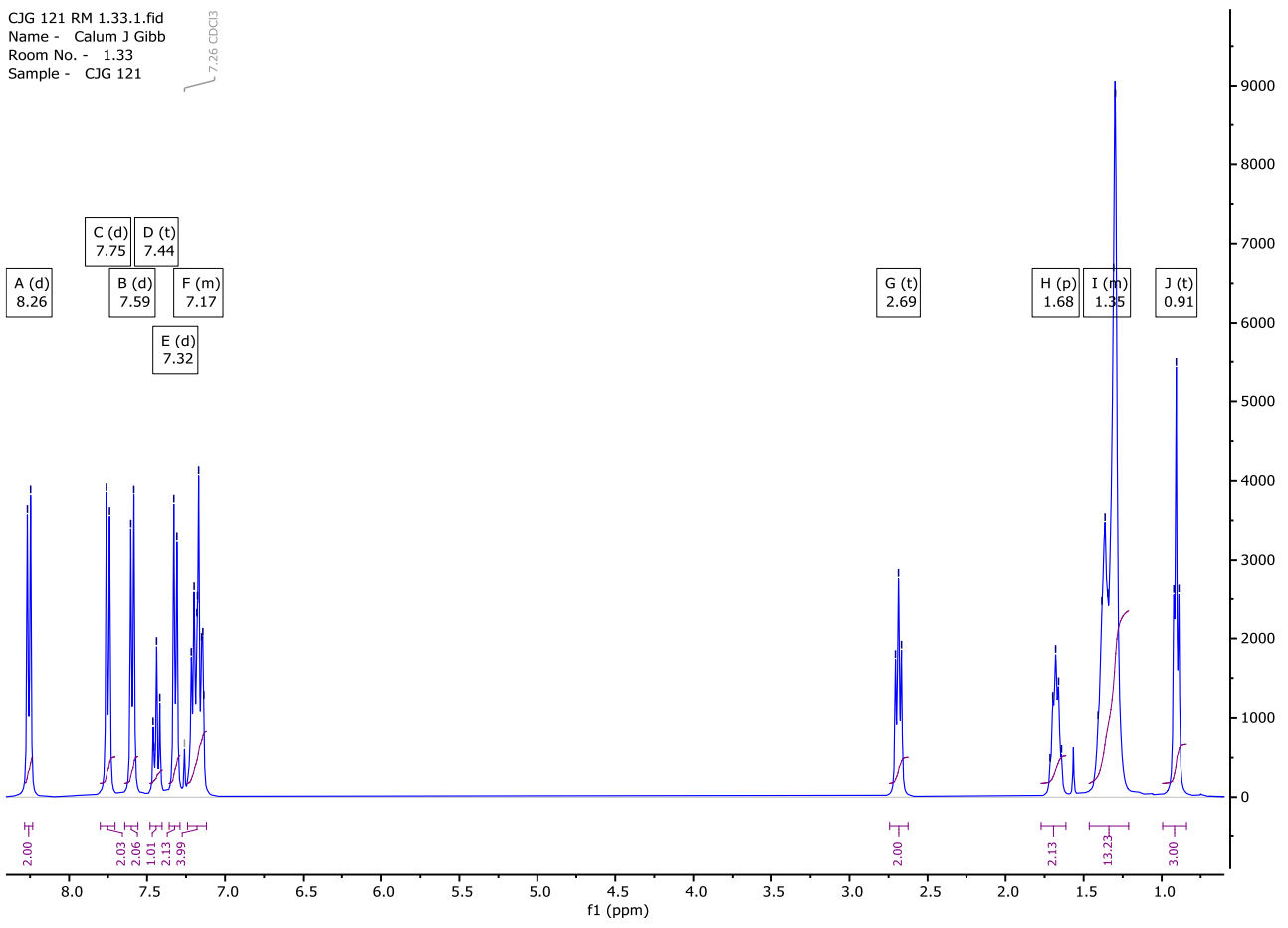

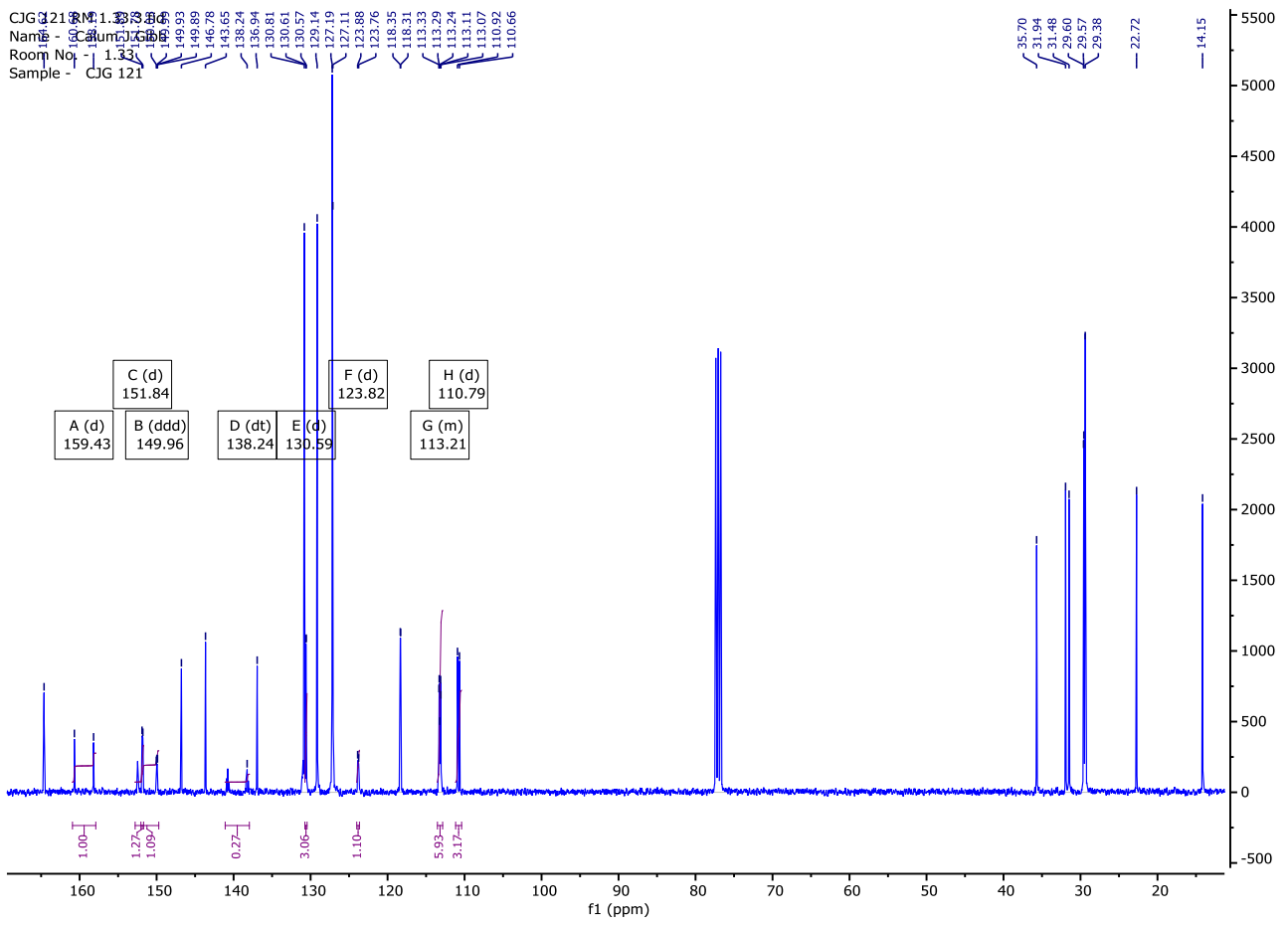

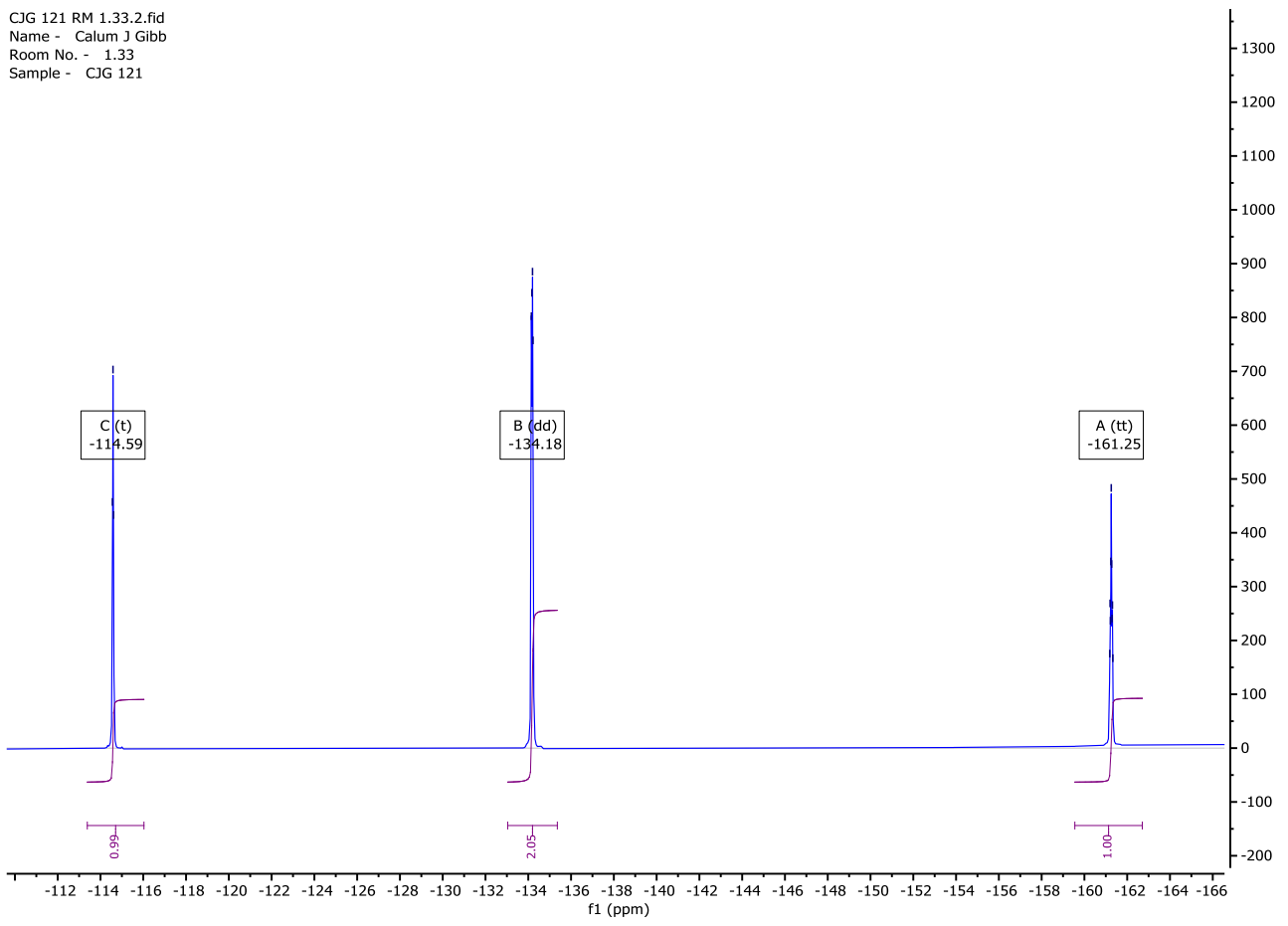

**Fig. S19.** $^{1}H$ [top], $^{13}C\{^{1}H\}$ [middle], and $^{19}F$ NMR spectra for **6** (**PPZGU-9-F**).

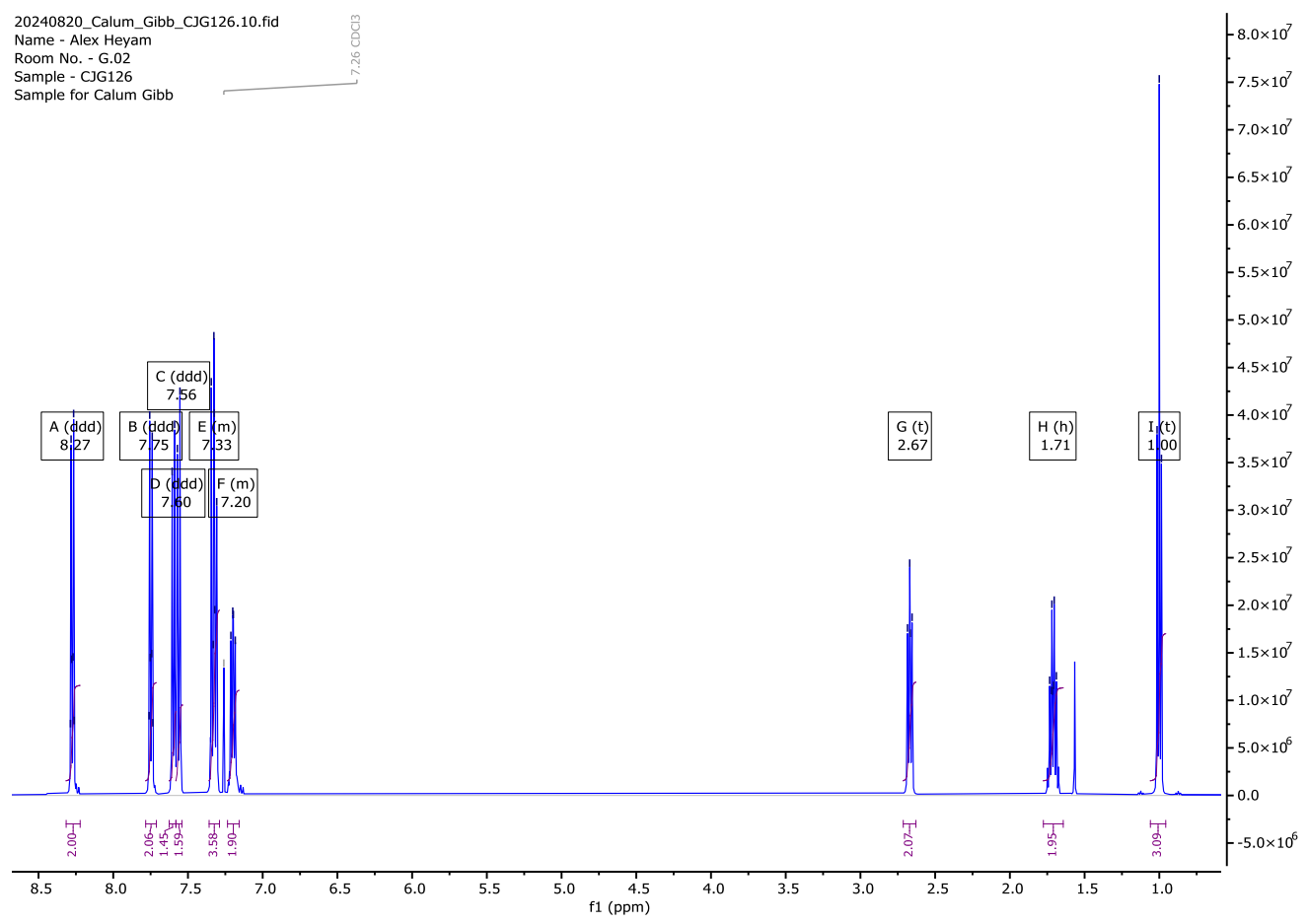

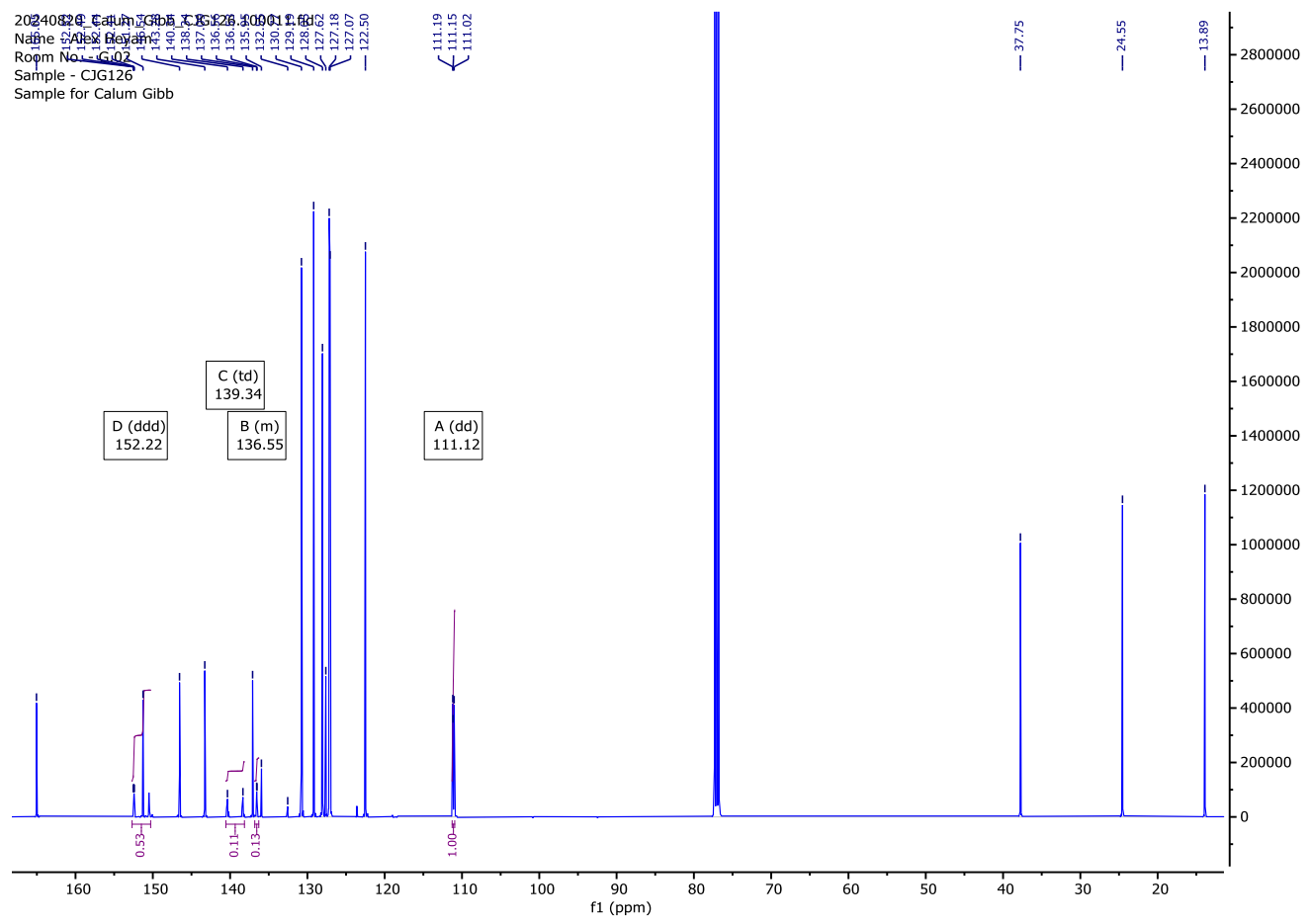

**Fig. S20.** $^{1}H$ [top], $^{13}C\{^{1}H\}$ [middle], and $^{19}F$ NMR spectra for **7** (**PPZPU-3-F**).

## 4 Supplementary References